\documentclass[aps,prb,twocolumn,showpacs,eqsecnum,superscriptaddress,floatfix,10pt]{revtex4-1}

\usepackage{graphicx}
\usepackage{epstopdf}
\usepackage{color}\usepackage{enumerate}
\usepackage{amsmath}\usepackage{amssymb}\usepackage{stmaryrd}\usepackage{wasysym}
\usepackage{multirow}
\usepackage{float}
\usepackage{longtable,booktabs}
\usepackage{diagrams}
\usepackage[unicode=true,pdfusetitle,
 bookmarks=false,colorlinks=true,citecolor=blue,urlcolor=blue,linkcolor=red]{hyperref}
\usepackage{url}
\usepackage{subfigure}%
\usepackage{dsfont}
\usepackage{amssymb}
\usepackage{extarrows}
\usepackage{mathtools}
\usepackage{xcolor}

\begin{document}

\title{From orbifolding conformal field theories to gauging topological phases}

\author{Xiao Chen}
\affiliation{Kavli Institute for Theoretical Physics, University of California at Santa Barbara, CA 93106, USA}
\affiliation{Department of Physics and Institute for Condensed Matter Theory, University of Illinois at Urbana-Champaign, 1110 West Green Street, Urbana, IL 61801-3080, USA}
\author{Abhishek Roy}
\affiliation{Institute of Theoretical Physics, University of Cologne, Zulpicher Strasse 77, D-50937 Cologne, Germany}
\author{Jeffrey C. Y. Teo}
\affiliation{Department of Physics, University of Virginia, Virginia 22904, USA}
\author{Shinsei Ryu}
\affiliation{
 James Franck Institute and Kadanoff Center for Theoretical Physics,
University of Chicago, Illinois 60637, USA
             }

\date{\today}

\begin{abstract} 
Topological phases of matter in (2+1) dimensions are commonly equipped with global symmetries, such as electric-magnetic duality in gauge theories and bilayer symmetry in fractional quantum Hall states. Gauging these symmetries into local dynamical ones is one way of obtaining exotic phases from conventional systems. We study this using the bulk-boundary correspondence and applying the orbifold construction to the (1+1) dimensional edge described by a conformal field theory (CFT). Our procedure puts twisted boundary conditions into the partition function, and predicts the fusion, spin and braiding behavior of anyonic excitations after gauging. We demonstrate this for the electric-magnetic self-dual $\mathbb{Z}_N$ gauge theory, the twofold symmetric $SU(3)_1$, and the $S_3$-symmetric $SO(8)_1$ Wess-Zumino-Witten theories.
\end{abstract}

\maketitle
\tableofcontents

\section{Introduction}

Topological phases of matter are those phases which defy the characterization 
in terms of spontaneous symmetry breaking, 
and hence cannot be described 
by the conventional Landau-Ginzburg theories.  
More precisely, it would be convenient to 
divide topological phases into two 
categories:
(1) short-range entangled symmetry-protected topological phases (SPT phase)~\cite{HasanKane10,QiZhangreview11,Senthil2014,RMP}, 
and 
(2) long-range entangled topologically ordered phases~\cite{Wilczekbook,Fradkinbook,Wenbook} 
Among others, one of the defining properties of phases 
with non-trivial topological order is the ground state degeneracy 
which depends on the topology of spatial manifolds.
This is the consequence of anyonic excitations (quasiparticles)
supported by the system. 
On the other hands, the ground state of SPT phases is unique on 
any spatial manifold.

In both kinds of topological phases, symmetries may play an important role.
In SPT phases such as those in Ref.~\onlinecite{KaneMele2D1, SchnyderRyuFurusakiLudwig08, Kitaevtable08, QiHughesZhang08, ChenGuLiuWen11, LuVishwanathE8, LevinGu12}, their very existence relies essentially on symmetries. Without symmetries, SPT phases may not be distinguishable from trivial phases. 
For topologically ordered phases, they can be enriched by the presence of symmetries.\cite{Essin2013, Mesaros2013, Lu2016, Lan_2016} For example, different topological liquids can be distinguished by symmetry fractionalization.\cite{BarkeshliBondersonChengWang14,TarantinoLindnerFidkowski15,Xie_Chen_2016} For topologically ordered phases, we can distinguish two kinds of
symmetries: (i) symmetries that can be realized locally and microscopically, and (ii)
those which can be realized at the level of low-energy (long-wave length), 
effective descriptions in terms of emergent anyonic excitations. 
The latter symmetry are often called anyonic symmetry.\cite{BaisSlingerlandCondensation, BarkeshliBondersonChengWang14, khan2014,TeoHughesFradkin15,TarantinoLindnerFidkowski15,Teotwistdefectreview}

In the presence of symmetries, it is interesting and important to discuss 
{\it defects} (more precisely, symmetry twist defects).\cite{Kitaev06, EtingofNikshychOstrik10,Bombin,YouWen,BarkeshliJianQi,KitaevKong12,TeoRoyXiao13long, khan2014,BarkeshliBondersonChengWang14,TeoHughesFradkin15,TarantinoLindnerFidkowski15, Teotwistdefectreview,Xie_Chen_2016} 
These defects are point-like objects in (2+1)-dimensions.
When excitations are (adabatically) transported around a symmetry twist 
defect and go back to their original locations, their anyon labels are rotated by the symmetry.
Defects may play an important role in the context of symmetry-breaking phases, 
e.g., pairing vortices in superconductors that associate with the $U(1)$ charge conservation symmetry breaking. 
Similarly, defects also play an important role in topological phases of matter.

In the context of SPT phases,  
the properties of symmetry twist defects can be used to 
diagnose properties of SPT phases.\cite{LevinGu12,Else_2014} 
More precisely, 
by gauging symmetry, i.e., by coupling the system to 
a dynamical gauge field,  
and promoting twist defects to dynamical anyonic excitations referred to as symmetry fluxes,
the properties of symmetry twist defects -- the braiding statistics 
obeyed by the dynamical symmetry fluxes -- 
can be used to diagnose and classify the properties of the original, ungagued SPT phases.\cite{LevinGu12, Ryu_2012, Sule_2013}

In the above procedure, gauging non-topologically ordered phases (SPT phases)
results in topologically ordered phases. 
On the other hands, one can gauge topologically ordered phases with global symmetry.
This procedure results in a new topologically ordered phases, which tend to have
more complicated topological order than the original topological phase.
\cite{BarkeshliBondersonChengWang14,TeoHughesFradkin15}
For example, systems with abelian topological order,
once gauged,
can lead to a new topologically ordered phase with
non-abelian topological order. 
Such topological phases are called twist topological liquids in
Ref.\ \onlinecite{TeoHughesFradkin15}. 

The bulk topological phase can be understood by studying the edge theory.\cite{Wen_1990} The bulk-boundary correspondence~\cite{FrohlichGabbiani90,MooreRead,ReadRezayi,Kitaev06}
states that the topological properties of
gapped (2+1)-dimensional topological phases can be extracted from 
their gapless edge theories, which can be described using 
(1+1)-dimensional conformal field theories (CFT).
In particular, 
superselection sectors of bulk quasiparticles are in one-to-one correspondence 
with primary fields in the edge CFT, and 
topological anyon properties such as
the bulk fusion rule, exchange statistics,
and monodromy braiding process, etc., all have corresponding 
descriptions in terms of the edge CFT
such as operator product expansions,  
conformal spin,
and modular $S$-transformations respectively.\cite{bigyellowbook}

Similarly, the bulk gauged topological liquid has a natural description
in terms of its edge theories, which is an {\it orbifold} CFTs.\cite{GINSPARG1988, DijkgraafVafaVerlindeVerlinde99} For a topological phase equipped with a global  symmetry, gauging symmetry in the bulk is equivalent to orbifolding symmetry in the CFT along the boundary.  In the context of (2+1)-dimensional SPT phases, 
orbifold CFTs have been used to diagnose SPT phases.
\cite{Ryu_2012, Sule_2013}
In this paper, we will discuss twist liquids from the boundary point of view. 
To gauge an (anyonic) symmetry in the bulk topological ordered phase,
we start from the edge CFT $\mathcal{C}$ that corresponds to the globally $\mathcal{G}$-symmetric topological state, and apply an orbifold operation, to obtain 
a new edge CFT $\mathcal{C}/\mathcal{G}$ that corresponds to the gauge symmetric twist liquid.
\begin{align}
\begin{diagram}
\stackrel{\mbox{Topological phase with}}{\mbox{global symmetry}}&\rTo^{\mbox{\small gauging}}&\mbox{Twist liquid.}\\\uTo\dTo&\stackrel{\mbox{bulk-boundary}}{\mbox{correspondence}}&\uTo\dTo\\\mbox{$(1+1)$d CFT $\mathcal{C}$}&\rTo_{\mbox{\small orbifolding}}&\mbox{orbifold CFT $\mathcal{C}/\mathcal{G}$}
\end{diagram}
\label{gaugingtransition}
\end{align}
This gives us a complementary view to the bulk description. Throughout this article, we focus on bosonic topological phases, whose local quasiparticles are all bosons. We speculate that a similar procedure may be carried out for fermionic phases by first promoting them into a bosonic ones through gauging the $\mathbb{Z}_2$ fermion parity symmetry. Certain concrete examples have been demonstrated in Ref.\ \onlinecite{KhanTeoHughesVishveshwara16}.

The structure of this paper is as follows. In Sec.~\ref{setup}, we first briefly review abelian topological phases and anyonic symmetries. Then we discuss the bulk-boundary correspondence and consider the general setup of orbifold CFTs. In Sec.~\ref{SPT_orbifold}, we study the orbifold CFT after gauging symmetries in SPT phases. In particular, we consider examples with $\mathbb{Z}_2$ symmetry and $\mathbb{Z}_3$ symmetry, and distinguish the resulting orbifolds between trivial and non-trivial parent SPTs. In Sec.~\ref{SET_orbifold}, we study the orbifold CFT after gauging anyonic symmetry in topological phases. We consider several examples explicitly and calculate their modular $\mathcal{S}$ and $\mathcal{T}$ matrices. 
They include gauging the $\mathbb{Z}_2$ electric-magnetic symmetry in the $D(\mathbb{Z}_N)$ quantum double, the $\mathbb{Z}_2$ bilayer (outer automorphism) symmetry of the $SU(3)$ Wess-Zumino-Witten (WZW) theory at level 1, and the $S_3$ triality symmetry of $SO(8)$ at level 1.
We summarize and conclude in Sec.~\ref{conclusion}.

%
%
%
%
%
%
%
%
%
%
%
\section{Generalities}
\label{setup}
Let us start by reviewing our ungauged theories 
(topologically ordered phases before gauging),
which are abelian topological phases in (2+1)d. 
In general, abelian topological phases in (2+1)d
can be understood pictorially 
as phases in which loop-like objects (Wilson loops or string-nets) 
are condensed.
The deformation of Wilson loops does not cost any energy. 
Of particular importance are
Wilson loops which wrap around non-contractible 
cycles of the spatial manifold -- they 
generate/measure 
the topological ground state degeneracy. 
The quasi-particle excitations (anyons) 
can be considered as the end of open strings (Wilson lines),
and hence
the types of quasi-particles depend on the types of strings. 
If a quasi-particle is dragged around another quasi-particle, 
the Wilson strings attached to them may intersect with each other. 
This intersection gives rise to a phase factor,
which is the braiding phase between the two quasi-particles. 

In a formal terms, the abelian topological phase can be described 
by a multi-component Chern-Simons theory (the K-matrix Chern-Simons theory), which is defined by the following Lagrangian (density): 
\begin{align}
\mathcal{L}_{CS}=
\frac{\epsilon^{\mu\nu\lambda}}{4\pi}
\boldsymbol\alpha_{\mu}^T\, \textbf{K}\,\partial^{\ }_{\nu}\boldsymbol\alpha^{\ }_{\lambda}
+\boldsymbol\alpha^T_{\mu}\, \vec{j}^{\mu}, 
\label{chern_simons}
\end{align}
where $\textbf{K}$ is an $N\times N$ symmetric matrix with integer entries, $\boldsymbol\alpha^T=(\alpha^1,\alpha^2,\ldots,\alpha^N)$ is the internal $U(1)^N$ gauge field coupled to the quasiparticle current $\vec{j}^{\mu}$. 

For this topological phase, by using a gauge invariance argument~\cite{Fradkinbook,Wenbook}, we can write down the effective edge theory,
\begin{align}
\mathcal{L}=\frac{1}{4\pi}\partial_t\boldsymbol\Phi^T(x) \textbf{K}\partial_x\boldsymbol\Phi(x)+\ldots
\end{align}
where $\boldsymbol\Phi$ is a $N$-component bosonic field. The quasi-particle excitaton on the boundary can be created by the vertex operator $e^{i\textbf{a}\cdot\boldsymbol\Phi}$.

Similar to the boundary, the bulk quasi-particle excitations $\psi^{\mathbf{a}}$ are labeled by
the $N$ component vector $\mathbf{a}$ living on the anyon integral lattice 
$\Gamma^*=\mathbb{Z}^N$.
They satisfy the Abelian fusion rule 
$\psi^{\mathbf{a}}\times\psi^{\mathbf{b}}
=\psi^{\mathbf{a}+\mathbf{b}}$. 
All the topological information of the quasi-particle are characterized by the $\textrm{K}$-matrix. 
The self and mutual braiding statistics for the quasi-particle are described by the $\mathcal{T}$ and $\mathcal{S}$ matrices.
For the K-matrix theory \eqref{chern_simons}, 
they are given by
$\mathcal{T}_{{\bf ab}}=\mathcal{T}_{{\bf aa}}\delta_{\bf ab}=e^{\pi i\textbf{a}^T\textbf{K}^{-1}\textbf{a}}$,
and 
$\mathcal{S}_{\bf ab}=e^{2\pi i\textbf{a}^T\textbf{\textbf{K}}^{-1}\textbf{b}}/\mathcal{D}$.
Here, $\mathcal{D}$ is the total quantum dimension,
$\mathcal{D}=\sqrt{|\det \textbf{K}|}$.\cite{Kitaev06}

On the anyon lattice, the states on the sublattice $\Gamma=\textbf{K}\mathbb{Z}^N$ are 
local particles.
They have trivial braiding statistics with all the other quasi-particles. Two quasi-particles that are differed by a local particle 
are topologically equivalent,
as they have the same braiding statistics. 
By only considering those quasi-particles living on the quotient lattice $\mathcal{A}=\Gamma^*/\Gamma=\mathbb{Z}^N/\textbf{K}\mathbb{Z}^N$,
we can remove this redundancy.

\subsection{anyonic symmetries}

When there is a unimodular (integral entries, unit determinant) 
matrix $W$ that leaves the K-matrix invariant, 
$W\textbf{K}W^T=\mathbf{K}$, 
we say $W$ is a symmetry operator for the topological phase.\cite{khan2014,Teotwistdefectreview}
These symmetry operators form the group of automorphisms,
\begin{align}
\mathrm{Aut}(\textbf{K})=\{W\in GL(N;\mathbb{Z})| W\textbf{K}W^T=\textbf{K}\}.
\end{align}
Aut($\textbf{K}$) is also the symmetry group of the anyon lattice and leaves the modular $\mathcal{S}$ and $\mathcal{T}$ matrices invariant.
On the other hand, 
when $\textbf{a}\to W\textbf{a}=\textbf{a}+\textbf{K}\mathbb{Z}^N$
for all $\textbf{a}$, 
$W$ does not change the quasi-particle class and only rearranges the local particle.
Such symmetry operators are trivial and form a normal subgroup called the inner automorphisms
\begin{align}
\mathrm{Inner}(\textbf{K})=\{W_0\in \mathrm{Aut}(\textbf{K})| W_0\textbf{a}=\textbf{a}+\textbf{K}\mathbb{Z}^N\}.
\end{align}
To enumerate non-trivial symmetry operators, 
we need to remove these trivial symmetry operators by quotienting and obtain the outer automorphisms
\begin{align}
\mathrm{Outer}(\textbf{K})=
\frac{\mathrm{Aut}(\textbf{K})}{\mathrm{Inner}(\textbf{K})}. 
\end{align}
The symmetry operations in Outer($\textbf{K}$) 
permute the quasi-particles on the anyon lattice.
We call such symmetry the anyonic symmetry.\cite{khan2014}

\begin{figure}
\centering
\includegraphics[width=2.95in]{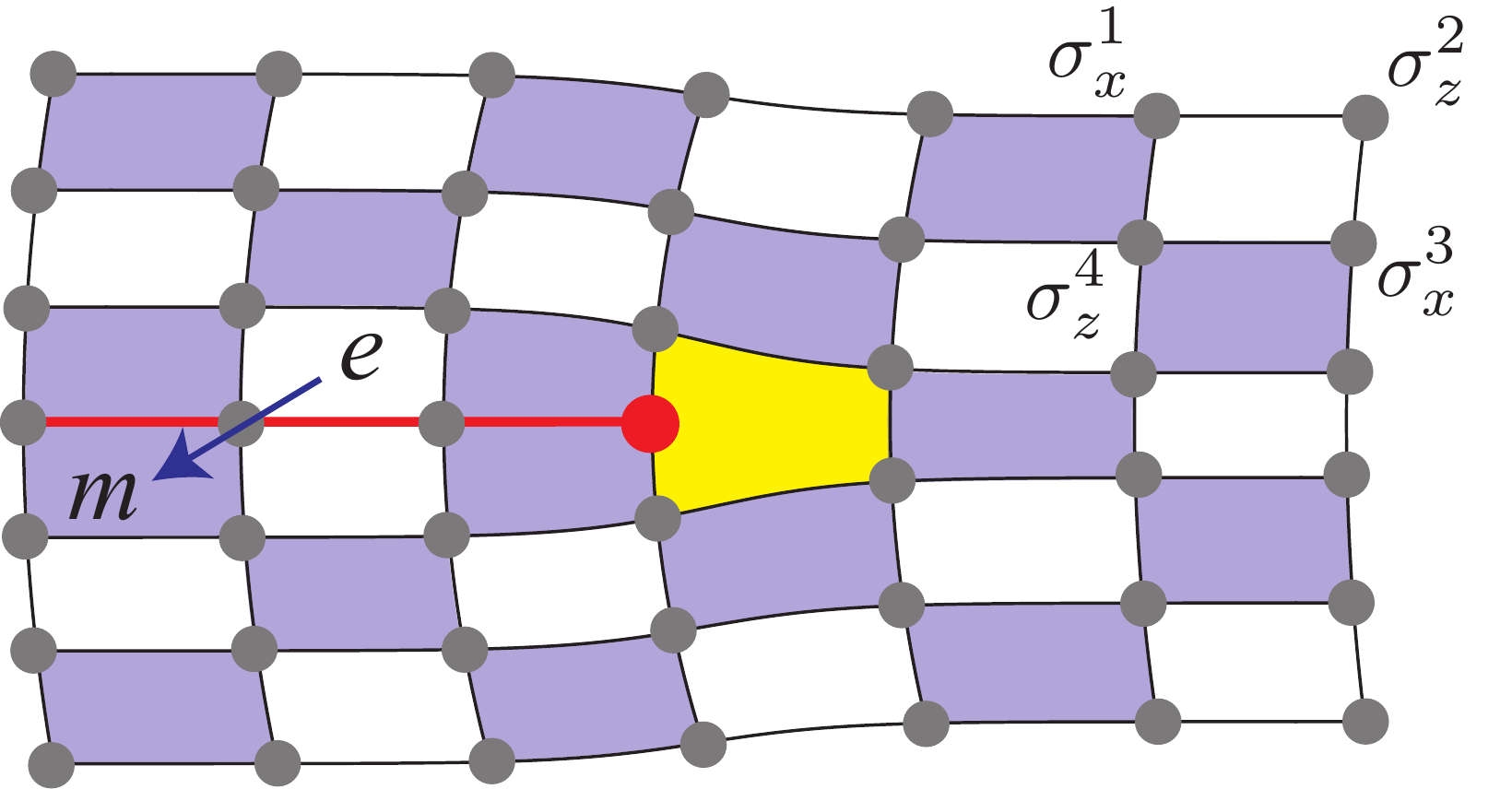}
\caption{Wen's plaquette lattice model with a dislocation in it.} 
\label{fig:Wen} 
\end{figure}

A good example is provided by 
the Kitaev toric code model,
which can be described by 
the effective field theory, 
the abelian Chern-Simons theory \eqref{chern_simons}
with ${\bf K}=2\sigma_x$.  
In the toric code model, 
the fundamental excitations are $\mathbb{Z}_2$ charge $e$ and flux $m$, 
which are bosons and obey mutual semionic statistics. 
$e$ and $m$ can combine together to form a fermionic 
composite quasi-particle $\psi=e\times m$. The $\mathcal{S}$ and $\mathcal{T}$ matrices are
\begin{align}
&\mathcal{S}=\frac{1}{2}\begin{pmatrix}
1 & 1 & 1 &1 \\
1 & 1 & -1 & -1 \\
1 & -1 & 1 & -1\\
1 & -1 & -1 & 1
\end{pmatrix},
              \nonumber\\
&\mathcal{T}=\mathrm{diag}(1,1,1,-1).
\end{align}
The toric code model has a global duality symmetry
which is an anyonic symmetry exchanging $e$ and $m$ quasi-particles, 
$e\leftrightarrow m$,
and leaves the $\mathcal{S}$ and $\mathcal{T}$ matrices invariant. 
This symmetry 
can be understood better by using Wen's plaquette lattice model with a bi-colored structure.\cite{Wenplaquettemodel} 
As shown in Fig.~\ref{fig:Wen}, this lattice model has spin $1/2$ degrees of freedom defined on the vertex of each plaquette,
and the physics for this model is equivalent to the toric code model. 
On this bi-colored lattice, the excitations $e$ and $m$ live on plaquettes with opposite colors.
The duality symmetry is implemented as swapping the colors of plaquettes globally. 
This corresponds to exchange the labels of charge $e$ and flux $m$, 
while leaving $\psi$ unchanged. 
Since the color of each plaquette is well defined, 
the duality symmetry is said to be weakly broken (according to Ref.\ \onlinecite{Kitaev06})
similar to the symmetry breaking phases.

\subsection{twist defects}

For a given aynonic symmetry $g$, one can introduce twist defects,
which are extrinsic classical defects in topological phases~\cite{Kitaev06,EtingofNikshychOstrik10,Bombin,YouWen,BarkeshliJianQi,KitaevKong12,TeoRoyXiao13long,khan2014,Teotwistdefectreview}.
When an abelian anyon ${\bf a}$ is dragged around 
a twist defect associated with an anyonic symmetry $g$, 
it will be rotated to $g{\bf a}$. 

Twist defects are non-abelian objects and their fusion and braiding statistics can be systematically described by the defect fusion category~\cite{EtingofNikshychOstrik10,TeoRoyXiao13long,BarkeshliBondersonChengWang14,TeoHughesFradkin15,Teotwistdefectreview}. In this paper, for the most part, we will focus on a parent abelian topological phase described 
by Eq.\ \eqref{chern_simons} equipped with anyonic symmetry group $\mathcal{G}$. 

When an abelian anyon ${\bf a}$ is dragged around 
a twist defect associated with an anyonic symmetry $g$, 
it will be rotated to $g{\bf a}$. 
Those anyons which are invariant under $g$ and satisfy ${\bf a}=g{\bf a}$,  
may be attached (fused) with the bare twist defect without an observable change. Other anyons with ${\bf a}\neq g{\bf a}$, fuse with the bare twist defect and form a composite which carries a \emph{species} label and lives in the quotient lattice $\mathcal{A}/(1-\mathcal{G})\mathcal{A}$ where $\mathcal{A}$ is the original anyon lattice.\cite{khan2014,Teotwistdefectreview}

Once again, a good example is provided by the toric code model. 
Twist defects in the toric code model 
are dislocations.\cite{YouWen, YouJianWen, Kitaev06, teo2013braiding}
There are two kinds of species labels and 
twist defects are denoted as $\sigma_0$ and $\sigma_1$. 
The fusion rules between them and the abelian anyon are
\begin{align}
\sigma_{\lambda}\times\psi=\sigma_{\lambda},\quad\sigma_{\lambda}\times e=\sigma_{\lambda}\times m=\sigma_{\lambda+1}
\end{align}
where $\lambda=0,1$ (mod 2) denotes the defect species. 
The fusion rules between twist defects are
\begin{align}
\sigma_0\times\sigma_0=\sigma_1\times\sigma_1=1+\psi,\quad\sigma_0\times\sigma_1=e+m
\end{align}

\subsection{gauging anyonic symmetries}
\label{gauge_anyonic}
An anyonic symmetry can be promoted to a local (gauge) symmetry;
we can gauge an abelian topologically-ordered phase by 
an anyonic symmetry $\mathcal{G}$ to generate a new topologically
ordered phase.\cite{BarkeshliBondersonChengWang14,TeoHughesFradkin15}
This can be achieved by 
proliferating twist defects and is analogous to $2+1$-dimensional melting transitions 
of ordered phases 
that restore broken symmetries by the vortex/defect proliferation. 
Once the phase transition happens, 
the parent Abelian phase
will be turned into 
a quantum mechanically more entangled topological phase with an increased topological entanglement entropy~\cite{KitaevPreskill06,BrownBartlettDohertyBarrett13} \begin{align}-S_0=\log\mathcal{D}_0\to-S=\log\mathcal{D}=-S_0+|\mathcal{G}|,\end{align} and the twist defect will become an intrinsic (non-abelian) quasi-particle excitation.

Before we discuss the twist liquid in detail, we will first briefly review quantum double models / discrete gauge theories, which are topological phases that arise from gauging a global symmetry in a trivial boson condensate. \cite{BaisDrielPropitius92, Bais-2007, Kitaev97, Propitius-1995, Preskilllecturenotes, Freedman-2004, Mochon04} 
For a finite discrete gauge group $G$, the anyon excitations are flux-charge composites labeled by the pair $\chi=([M],\rho)$. The flux component $M$ is characterized by its conjugacy class in the gauge group $G$,
\begin{align}
[M]=\{ M^\prime \in G:  M^\prime= N M N^{-1} \mbox{for some}\ N\in G \}
\end{align} 
The charge component is an irreducible representation $\rho$ of the centralizer of $M$
\begin{align}
Z_M=\{ N\in G: NM=MN  \}.
\end{align}
For the pure charge excitation with $[M]=[1]$, $\rho$ is the irreducible representation of the whole group since $Z_M=G$.

The quasi-particle structure in the twist liquid is a generalization of the quantum double model and the quasi-particle excitations have three components: flux, super-sector and charge, which can be labeled by the 3-tuple $\chi=([M],\boldsymbol\lambda, \rho)$.\cite{TeoHughesFradkin15} Here $[M]$ is still the conjugacy class for the gauged $G$-symmetry. $\boldsymbol\lambda$ is a super-sector of  species labels drawn from the quotient group $\mathcal{A}_M=\mathcal{A}/(1-M)\mathcal{A}$. The anyonic symmetry (defined by the conjugacy class $[M]$) can permute elements living on the quotient lattice $\mathcal{A}_M$ and therefore force them to combine together and form into super-sector
\begin{align}
\boldsymbol\lambda=\lambda_1+\ldots+\lambda_l,\quad \lambda_i\in\mathcal{A}_M
\end{align}
where the set of species labels in $\boldsymbol\lambda$ are permuted by the elements in the centralizer, i.e., $\lambda_j=N\lambda_i$ for some $N\in Z_M$.

The last element in the 3-tuple is characterized by the irreducible representation of the restricted centralizer of M, which satisfies
\begin{align}
Z_M^{\boldsymbol\lambda}=\{ N\in Z_M: N\lambda_1=\lambda_1  \}
\end{align}
This means that $Z_M^{\boldsymbol\lambda}$ is the subgroup in the centralizer $Z_M$ that fixes a particular choice of speices  $\lambda_1$ in $\boldsymbol\lambda$. Here we arbitrarily choose $\lambda_1$ for convenience.  

If the anyonic symmetry is abelian, the quasi-particle structure in the twist liquid is much simpler. This is because the conjugacy class is each element in the group $G$ and the centralizer for arbitray $[M]$ is the whole group. For instance, in the toric code model with four quasi-particle excitations $I, e, m, \psi$, there is an electric-magnetic duality symmetry which exchanges $e\leftrightarrow m$ and keeps $1$ and $\psi$ invariant. This is a $\mathbb{Z}_2$ anyonic symmetry with $G=\{1,\sigma\}$. For $[1]$ (zero flux sector), after gauging  anyonic symmetry, the allowed species labels are $\boldsymbol\lambda=I,e+m,\psi$, where $e$ and $m$  change to each other under $\sigma$ and need to combine into $e+m$. The charge component is determined by the restricted centralizer group and is equal to
\begin{align}
Z_{[1]}^I=\mathbb{Z}_2,\quad Z_{[1]}^{e+m}=0,\quad Z_{[1]}^{\psi}=\mathbb{Z}_2
\end{align}
Notice that $ Z_{[1]}^{e+m}=0$ is trivial here and consists only identity element. In the $[\sigma]$ flux sector, there are only two species labels $\{0, 1\}$, where we denote $0$ for $I$ or $\psi$ and $1$ for $e$ or $m$. The charge component satisfies
\begin{align}
Z_{[\sigma]}^0=\mathbb{Z}_2,\quad Z_{[\sigma]}^1=\mathbb{Z}_2
\end{align}
In total, there are nine excitations
\begin{align}
&1=([1],1,\rho_+),\quad z=([1],1,\rho_-)\nonumber\\
&\mathcal{E}=([1],e+m,\rho_0)\nonumber\\
&\psi=([\sigma],1,\rho_+),\quad \bar{\psi}=([\sigma],1,\rho_-)\nonumber\\
&\sigma=([\sigma],1,\rho_+),\quad \sigma^\prime=([\sigma],1,\rho_-)\nonumber\\
&\bar{\sigma}=([\sigma],e,\rho_+),\quad \bar{\sigma}^\prime=([\sigma],e,\rho_-)
\end{align}
We can divide them into three types of excitations: (1) $1$, $z$; (2) $\mathcal{E}$ and (3) $\psi,\bar{\psi},\sigma,\sigma^\prime,\bar{\sigma},\bar{\sigma}^\prime$. Type (1) has zero flux and can be understood as  anyon ${\bf a}$ in $\mathcal{A}$ coupled with gauge charge. ${\bf a}$ needs to be invariant under anyonic symmetry and the composite particle is an abelian excitation in the twist liquid. Type (2) still has zero gauge flux but is a non-abelian excitation.  This is because the super-sector is formed by multiple $\lambda_i$, i.e., $\boldsymbol\lambda=\sum_i\lambda_i$. They need to group in this way so that $\boldsymbol\lambda$ is invariant under anyonic symmetry. Type (3) is the most interesting excitation. It has non-trivial gauge flux and corresponds to the non-abelian twist defect before gauging. They can carry species labels and couple with gauge charge to form a composite excitation. These three types of excitations are general in the twist liquid with abelian group $G$. Later on, we will use bulk-boundary correspondence and explicitly show how to construct these three types of characters in the orbifold CFT.


Finally, the gauging procedure can be also understood as
the reverse of the condensation 
of bosons,
\begin{align}
\parbox{9em}{Topological phase with global anyonic symmetry}
\xrightleftharpoons[\mbox{Condensation}]{\mbox{Gauging}} \mbox{Twist liquid}
\end{align}
This is referred to as anyon condensation in the literature.\cite{BaisSlingerlandCondensation,Kong14} After condensation, the twist liquid
is turned into the parent abelian phase with global symmetry. 


\subsection{edge theories}

For $(2+1)$d topological phases, 
physics at their boundary (edge) 
is described by 
corresponding $(1+1)$d conformal field theories (CFTs).\cite{Wen_1990, MooreRead,ReadRezayi,Kitaev06}
In this paper, 
we will make use of this bulk-boundary correspondence to study
topological order before and after gauging anyonic symmetry.\cite{GINSPARG1988, DijkgraafVafaVerlindeVerlinde99, Moore_Seiberg_1989}
In particular, topological order which results by gauging a parent phase
can be studied by using orbifold CFTs (see below). 


The bulk-boundary correspondence asserts that 
there is a one-to-one correspondence between 
quasi-particle excitations (anyons) in the bulk 
and primary fields living on the boundary. 
One way to understand this is to note that 
bulk quasi-particle excitations 
are emergent collective objects, and hence, 
unlike fundamental particles (electrons), 
bear certain kinds of non-locality.
In particular, 
exciting a quasi-particle will create a string/branch cut emanating from it,
which may create a twist on the boundary condition. 
In the language of the edge CFT, this twist is generated by the primary field. 

The lowest energy state (heighest weight state in the Virasoro algebra) 
in the presence of this twist 
corresponds to a quasi-particle state in the bulk.
Furthermore, for the tower of states built on the lowest energy state
one can define a character $\chi(\tau)$. 
More precisely, 
let us consider a $(2+1)$d topological phase defined on the disk, 
which supports, as its boundary, a circle $S^1$.
Including the (imaginary) time direction,
which is also a circle $S^1$, 
the edge CFT lives on a spacetime torus $S^1\times S^1$. 
Using the Hamiltonian $H_0$ and the momentum $P_0$ of the CFT, 
the character is then defined by
\begin{align}
\chi_j(\tau) =\mathrm{Tr}_{O_j}\left[e^{2\pi i\tau_1 P_0-2\pi \tau_2 H_0}\right]
\label{chars}
\end{align}
where $O_j$ denotes a primary field (state) 
and the trace is taken over all states 
built on the primary state; 
$\tau=\tau_1+i\tau_2$ is the modular parameter of the torus,
which parameterizes distinct conformally flat torii. Here $\tau_1$ is the spatial period, and $\tau_2$ is the temporal period.

Different modular parameters, 
when related to each other by modular transformations $PSL(2,\mathbb{Z})$
represent an equivalent torus. 
In other words, the modular transformations are  
large coordinate or large diffeomorphism transformations,
which leave a torus invariant but may change the modular parameter $\tau$.
The modular transformations are generated by the so called Dehn twists $\mathcal{T}:\tau\to\tau+1$ and $\mathcal{S}$ which flips $\tau_1$ and $\tau_2$: $\tau\to-1/\tau$. 
The modular transformations
act covariantly on the characters as well. 
Under the $\mathcal{T}$ and $\mathcal{S}$ modular transformation,
$\chi_j(\tau)$ transforms as:
\begin{align}
\nonumber \chi_j(\tau+1)&=e^{2\pi i(h_j-\frac{c}{24})}\chi_j(\tau)=\mathcal{T}_{jj}\chi_j(\tau),
\\
\chi_j(-{1}/{\tau})&=\sum_{j^{\prime}}\mathcal{S}_{jj^{\prime}}\chi_{j^{\prime}}(\tau), 
\end{align}
where $\mathcal{T}$ is taken as a diagonal matrix and 
describes the conformal dimension $h$ for the corresponding primary field $O_j$ of each character,
whereas $c=c_R-c_L$ is the chiral central charge for the total system. 

The $\mathcal{T}$ and $\mathcal{S}$ matrices defined here for the characters 
are closely related to 
the $\mathcal{T}$ and $\mathcal{S}$ transformations of the degenerate ground state Hilbert space when the $(2+1)d$ topological bulk is put on a torus.\cite{Wilczekbook,Fradkinbook,Wenbook} 
The bulk and boundary $\mathcal{T}$
differ only by a phase factor 
$2\pi c/24$;
in the following calculations for the boundary theory, 
we will neglect this overall phase. 
As for the $\mathcal{S}$ matrix, 
the bulk $\mathcal{S}$ matrix is the same as 
the edge $\mathcal{S}$ matrix. 
Thus, the boundary
$\mathcal{T}$ and $\mathcal{S}$ 
encodes the same topological information as their bulk counterparts. 
For example, 
the fusion rule for bulk-quasiparticles 
(and for boundary primary fields)
can be read off from $\mathcal{S}$ by using the Verlinde formula,\cite{Verlinde88}
\begin{align}
N^x_{yz}=\sum_w\frac{S_{xw}S_{yw}S^*_{zw}}{S_{1w}},
\label{verlinde}
\end{align}
where $N^x_{yz}$ is the fusion matrices that characterize fusion rules $x\times y=\sum_zN^z_{xy}z$.

\subsection{symmetry operator and orbifold CFT}

By the bulk-boundary correspondence, 
there is a description of gauged twist topological liquid
in terms of edge CFTs. 
As discussed before, (unparied) twist defects in the bulk will leave a twist on the boundary. 
Once the symmetry $\mathcal{G}$ is gauged in the bulk,
this corresponds to promotion of twist defects as dynamical quantum excitations.
This means, in the edge CFT, 
that we need to consider all possible twisted spatial boundary conditions. 
Moreover, after gauging the symmetry $\mathcal{G}$, 
the Hilbert space of the edge CFT is restricted to the $\mathcal{G}$-invariant subspace, 
which can be realized by applying the projection operator $\hat{P}$
on the Hilbert space. 
In short, 
boundary CFTs corresponding to bulk gauged twist topological liquid
are orbifold CFTs by symmetry $\mathcal{G}$.

Here we will briefly review the construction of orbifold CFTs.
For details, see Ref.\ \onlinecite{bigyellowbook,GINSPARG1988, DijkgraafVafaVerlindeVerlinde99, Moore_Seiberg_1989}. 
Consider a rational CFT $\mathcal{C}$ with a discrete symmetry group $\mathcal{G}$.
An orbifold CFT $\mathcal{C}/\mathcal{G}$ is constructed by moding out the group $\mathcal{G}$ and the Hilbert space is projected on to the $\mathcal{G}$ invariant subspace. 
The projection operator
\begin{align}
P=\frac{1}{|\mathcal{G}|}\sum_{g\in \mathcal{G}}g
\label{proj}
\end{align}
is inserted in the trace in Eq.\ \eqref{chars}. 
In the spacetime (path-integral) picture, 
the operator $g$ 
in the projector can be interpreted as twisting the boundary condition in the time direction. 
In addition to the projector \eqref{proj}, we will also need to consider 
a more generic form of the projectors
$P=\frac{1}{|\mathcal{G}|}\sum_{g\in \mathcal{G}}\varepsilon(g)g$,
which projects on the different sectors of the Hilbert space,
where $\varepsilon(g)$ is a $g$-dependent phase factor.

Under the $\mathcal{S}$ transformation, 
the twist in the time direction will become a twist in the spatial direction.
We therefore need to included twisted characters
in the presence of a spatial twist boundary condition by $h\in \mathcal{G}$. 
For each twisted character, we consider, as before, the projection operator
$P_n=\frac{1}{|\mathcal{G}|}\sum_{g\in \mathcal{G}}\varepsilon_n(h,g)g$. The character under this projection is a combination of partition functions with twisted boundary conditions.\cite{Ginsbarg_1991} For instance, if $\mathcal{G}$ is an abelian $\mathbb{Z}_N$ symmetry, $P_n$ is simply written as $P_n=\sum_{k=0}^{N-1}\omega ^{-nk}g^k/N$ where $\omega=e^{-2\pi i /N}$ and $g$ is the generator of $\mathbb{Z}_N$. The character is $\chi_n^h=\sum_{k}^{N-1}\omega^{-nk}Z(h,g^k)/N$.

All together, we need to consider 
the set of partition functions $Z(g,h)$ in the presence of 
twisted boundary condition by $g$ ($h$) in the temporal (spatial) direction.  
In general, 
these partition functions transform 
under the $\mathcal{S}$ and 
$\mathcal{T}$ transformation, 
as
\begin{align}
&Z(g,h)\overset{\mathcal{S}}{\longrightarrow}Z(h,g),
\nonumber\\
&Z(g,h)\overset{\mathcal{T}}{\longrightarrow}Z(gh,h). 
\label{modu_tran}
\end{align}
up to some phases.
Note that when $\mathcal{G}$ is a non-abelian group, we require that for $Z(g,h)$, $g$ and $h$ must commute with each other.

Combining these $Z(g,h)$ together, we construct the characters for the orbifold CFT as
\begin{align}
\chi^h_n&=\frac{1}{|Z_h|}\sum_{g\in Z_h}\epsilon_n(g,h)Z(h,g), 
\label{char_def}
\end{align}
where $Z_h$ is the centralizer subgroup of $h$ containing all $g$ that commutes with $h$. The characters are the partition functions in the $h$-twisted Hilbert space
under the projection operator $P$. 
Using Eq.\ \eqref{modu_tran}, 
we can show that the characters of the orbifold CFT form a complete basis under the modular transformations.
Moreover, the chiral central charge of the orbifold CFT is the same as the original CFT since the overall phase under $\mathcal{T}$ transformation does not change.


The above framework is the general structure of orbifold CFT. Eq.\eqref{char_def} can be directly applied to the edge theory of a topological phase after gauging global symmetry in SPT phase, where various twists are introduced on the boundary CFT. Furthermore, we can generalize this method to the more complicated orbifold CFT after gauging anyonic symmetry $G$ in topological phase. Before we investigate them, we first discuss the general construction of the edge theory for abelian topological phase in the next section.

\subsection{Abelian topological phases and edge theories}
\label{edge_cft}

The effective edge theory 
of an abelian topological phase 
can be described by the $N$-component bosonic field theory:
\begin{align}
\mathcal{L}=\frac{1}{4\pi}\left(\partial_t\vec{\Phi}^T(x) \textbf{K}\partial_x\vec{\Phi}(x)-\partial_x\vec{\Phi}^T(x) \textbf{V}\partial_x\vec{\Phi}(x)\right), 
\label{lutt}
\end{align}
where the bosonic field $\vec{\Phi}(x)$ is 
a compact variable
\begin{align}
\vec{\Phi}(x)\equiv\vec{\Phi}(x)+2\pi\vec{n},
\end{align}
and 
$\textbf{V}$ is a symmetric and positive definite matrix that accounts for the interaction on the edge.
Unlike the K-matrix, the information encoded 
in $\mathbf{V}$ is non-universal.

Below we will canonically quantize this theory and write down all the characters explicitly. 
Here we follow the method used in Ref.\ \onlinecite{Hsieh2014}.
For each component of the bosonic field 
$\vec{\Phi}$, 
the canonical commutation relation is given 
by
\begin{align}
[\phi^I(x),\partial_x\phi^J(x^{\prime})]=2\pi i(\textbf{K}^{-1})^{IJ}\delta(x-x^{\prime})
\end{align}
The K-matrix can be diagonalized and written as $\textbf{K}=U^{T}\eta U$, 
where $\eta$ is a signature matrix 
with $\pm 1$ in its diagonal entries. 
We can define a new multi-component boson operator $\vec{\varphi}(x)=U\vec{\Phi}(x)$ so that 
Eq.\ \eqref{lutt} takes the form
\begin{align}
\mathcal{L}=\frac{1}{4\pi}\left(\partial_t\vec{\varphi}^T(x) \eta\vec{\varphi}(x)-\partial_x\vec{\varphi}^T(x) \partial_x\vec{\varphi}(x)\right), 
\label{diag}
\end{align}
where we assume $(U^{-1})^T\textbf{V}U^{-1}=I$ for simplicity.
The new boson operator $\vec{\varphi}(x)$ satisfies
\begin{align}
\vec{\varphi}(x)\equiv\vec{\varphi}(x)+2\pi U\vec{n}
\label{comp}
\end{align}
Each component satisfies the commutation relation:
\begin{align}
[\varphi^I(x),\partial_x\varphi^J(x^{\prime})]=2\pi i\eta^{IJ}\delta(x-x^{\prime})
\end{align}
The mode expansion for $\vec{\varphi}(x)$ is given by
\begin{align}
\varphi^I(t,x)&=\varphi_0^I-(t-\eta^{II}x)p^I
\nonumber \\
&\quad 
+i\sum_{n\neq 0} b_n^Ie^{-in(t-\eta^{II}x)}, 
\label{mode}
\end{align}
where $[\varphi_0^I,p^I]=i\delta^{IJ}$ and $[b_n^I,b_m^J]=\frac{1}{m}\delta^{IJ}\delta_{n+m}$.
The Hamiltonian and the total momentum are
\begin{align}
\nonumber H_0&=\frac{1}{4\pi}\int_0^{2\pi}dx\,
\partial_x\vec{\varphi}^T(x)\partial_x\vec{\varphi}(x)\\
&=\frac{1}{2}\vec{p}^T\vec{p}
-\frac{1}{24}\mathrm{tr}\,(\eta\eta)+\sum_{n=1}^{\infty}n^2\vec{b}_{-n}^T\vec{b}_n,
\nonumber \\
P_0&=\frac{1}{4\pi}\int_0^{2\pi}dx\, \partial_x\vec{\varphi}^T(x)\eta\partial_x\vec{\varphi}(x)
\nonumber \\
&=\frac{1}{2}\vec{p}^T\eta\vec{p}-\frac{1}{24}\mathrm{tr}(\eta)+\sum_{n=1}^{\infty}n^2\vec{b}_{-n}^T\eta\vec{b}_n
\end{align}
where $\mbox{Tr}(\eta)$gives the chiral central charge. Since $\vec{p}$ is the momentum conjugate 
to the zero modes $\vec{\varphi}_0$,
the compactification condition of
$\vec{\varphi}(x)$ 
leads to the quantization condition for $\vec{p}$: 
$\vec{p}=(U^{-1})^T\vec{m}$ where $m^I\in \mathbb{Z}$. 
Notice that $\vec{m}$ can be written as $\vec{m}=\textbf{K}\vec{\Lambda}+\vec{\lambda}$, 
where $\vec{\Lambda}$ and $\vec{\lambda}$ 
are integer valued vectors.  
Here, 
$\vec{\lambda}$ lives in the unit cell of the anyon lattice $\Gamma^\ast$ and characterizes different twist
boundary conditions for $\vec{\varphi}(t,x)$.
On the other hand,
for $\vec{\lambda}^T=0$, 
$\vec{p}=\eta U\vec{m}$, 
this corresponds to the untwisted boundary condition. 
Physically, 
$\vec{\lambda}$'s 
represent 
different bulk excitations in the (2+1)d topological phase;
When there is a quasi-particle excitation in the bulk, 
the edge theory is subject to  
the corresponding twist boundary condition. 

The partition functions of the edge theory 
on the spacetime torus
under the twist boundary conditions 
form the characters.
In total, there are $|\det(\textbf{K})|$ characters. 
The partition function for each character $[{\lambda}]$ is given by
\begin{align}
\chi_{\lambda}(\tau)=\textrm{Tr}_{\lambda}\left[e^{2\pi i\tau_1P_0}e^{-2\pi\tau_2 H_0}\right]
\end{align}
where $\tau=\tau_1+i\tau_2$.
Under the ${T}$ transformation,
\begin{align}
\chi_{\lambda}(\tau+1)=
e^{2\pi i\vec{\lambda}^T\textbf{K}^{-1}\vec{\lambda}/2}
\chi_{\lambda}(\tau). 
\end{align} 
where we neglect the overall phase $-2\pi c/24$. 
Under the $\mathcal{S}$ transformation,
\begin{align}
\chi_{\lambda}(-1/\tau)=\sum_{\lambda^{\prime}} \frac{1}{\sqrt{|\det(\textbf{K})|}}
e^{-2\pi i
\vec{\lambda}^T
\textbf{K}^{-1}
\vec{\lambda}^{\prime}}
\chi_{\lambda^{\prime}}(\tau). 
\end{align}

When $\eta=I_{N\times N}$, i.e., the bosonic theory is chiral, 
the character takes the following form:
\begin{align}
\chi_{\lambda}(q)=\frac{1}{\eta(\tau)^N}\sum_{\vec{\Lambda}}q^{\frac{1}{2}(\textbf{K}\vec{\Lambda}+\vec{\lambda})^T\textbf{K}^{-1}(\textbf{K}\vec{\Lambda}+\vec{\lambda})}
\end{align}
where $q=e^{i2\pi\tau}$ and $\eta(\tau)$ is the Dedekind eta function
\begin{align}
\eta(\tau)=q^{\frac{1}{24}}\prod_{n=1}^{\infty}(1-q^n). 
\label{eta_func}
\end{align}

When $N=1$, the K-matrix is just
an integer $K$. This is the $U(1)_{K/2}$ theory. 
Here we only consider when $K$ is even number. 
The character takes the form
\begin{align}
\nonumber\chi_{\lambda}(\tau)&=\frac{1}{\eta(\tau)}\sum_n q^{\frac{K}{2}(n+\frac{\lambda}{K})^2}
=\frac{1}{\eta(\tau)}\Theta_{\lambda,K}(q)
\label{theta_f}
\end{align}
where $0\leq\lambda<K$. 
Under the ${T}$ transformation
\begin{align}
\Theta_{\lambda,K}(\tau+1)=e^{2\pi i\frac{\lambda^2}{2K}}\Theta_{\lambda,K}(\tau). 
\end{align}
On the other hand, under the ${S}$ transformation
\begin{align}
\Theta_{\lambda,K}(-{1}/{\tau})=\sqrt{\frac{-i\tau}{K}}\sum_{n}\Theta_{\lambda^{\prime},K}(\tau)e^{-2\pi i\frac{\lambda^{\prime}\lambda}{K}}.
\end{align}

\section{SPT phases}
\label{SPT_orbifold}

To warm up, 
we shall first consider several simple examples of symmetry-protected topological phases (SPT) 
in $(2+1)$ dimensions.\cite{HasanKane10,QiZhangreview11,Senthil2014,RMP}
(Recall that SPT phases are short-range entangled states and do not have topological order.) 
On the boundary, there are gapless degrees of freedom
protected by the symmetry $\mathcal{G}$. 
The examples we consider are SPT phase protected by global $\mathbb{Z}_K$ symmetry.\cite{ChenGuLiuWen12,LuVishwanathE8, LevinGu12} 
We will calculate the edge partition function with $\mathbb{Z}_K$ orbifold. 
This part of calculation actually has been done in our previous paper and the detail can be found there (Ref.\ \onlinecite{Sule_2013}). 
We briefly review this calculation here for two reasons.
First, later in Sec. \ref{orb_z2_spt} and \ref{orb_z3_spt}, we will analyze $K=2$ and $K=3$ in detail. 
Second, this is a simple case of orbifold CFT. 
All the other orbifold CFTs related with gauging the anyonic symmetry discussed later in this paper are based on the same method.

Here we study the edge CFT which can be described by the Luttinger liquid in
Eq.\, \eqref{lutt} with $\det(\textbf{K})=1$ and protected by a discrete symmetry $\mathcal{G}$. 
After gauging the symmetry $\mathcal{G}$, the SPT phase will be promoted to a topological phase and on the boundary, the effective edge theory becomes an orbifold CFT. By studying the characters for this orbifold CFT and calculating the modular $\mathcal{T}$ and $\mathcal{S}$ matrices, we can extract topological information of the gauged topological phase in the bulk.

The gapless edge state for a non-trivial SPT phase cannot be gapped out without breaking symmetry and multiple copies of SPT are required to gap out the edge without breaking the symmetry. 
It is argued that the CFT after orbifolding symmetry $\mathcal{G}$ is {\em anomalous}. The partition function for this orbifold CFT is not invariant under modular transformation and therefore cannot be realized 
as an isolated $(1+1)$-dimensional system. 
At least two copies of orbifold CFT are needed to construct the modular invariant partition function and this construction can be understood through the boson condensation mechanism.\cite{BaisSlingerlandCondensation} 
On the other hand, for a trivial SPT, the edge gapless state can be gapped out by adding an interaction term without breaking symmetry $\mathcal{G}$. The edge orbifold CFT after gauging is anomaly-free and the partition function is modular invariant. 
%

The edge of the $\mathbb{Z}_K$ SPT phase can be described by 
the two-component K-matrix theory with $\textbf{K}=\sigma_x$. 
This edge theory is protected by $\mathcal{G}=\mathbb{Z}_K$ symmetry, 
which acts on the boson fields as 
\begin{align}
\phi^1\to \phi^1+2\pi k/K,\quad \phi^2\to\phi^2+2\pi kq/K
\label{z2_symm}
\end{align}
where $k=0,1,2, ..., K-1$ represents an element inside the symmetry group and $q=0,1,2, ..., K-1$ is fixed by SPT phase.

As shown in Eq.\ \eqref{diag}, we can diagonalize the K-matrix 
by introducing $\vec{\varphi}=(\varphi^1,\varphi^2)^T$ as
\begin{align}
\begin{pmatrix}\varphi^1\\ \varphi^2\end{pmatrix}=\sqrt{\frac{1}{2}}\begin{pmatrix}1&1\\1&-1\end{pmatrix}\begin{pmatrix}\phi^1\\ \phi^2\end{pmatrix}.
\end{align}
The new bosonic fields obey the compactification condition:
\begin{align}
\nonumber \varphi^1(t,x)&\equiv\varphi^1(t,x)+\frac{2\pi}{\sqrt{2}}(n_1+n_2),
\\
\varphi^2(t,x)&\equiv\varphi^2(t,x)+\frac{2\pi}{\sqrt{2}}(n_1-n_2),
\end{align}
with $p^1=\frac{1}{\sqrt{2}}(n_1+n_2)$ and $p^2=\frac{1}{\sqrt{2}}(n_1-n_2)$.

Once $\mathbb{Z}_K$ symmetry is gauged, 
$\varphi^2$ will be subjected to twisted boundary conditions 
in spatial and time directions. This twist will be reflected in the partition function. 
There are $K^2$ sectors of partition function $Z^{k,l}$, 
where $k,l=0,1,2,\ldots, K-1$ represent the boundary conditions in spatial and time directions. The twist boundary condition for $\varphi^1$ field in the spatial direction will lead to the modified quantization condition for $p^1$ and $p^2$, 
\begin{align}
&p^1=\frac{1}{\sqrt{2}}(n_1+\frac{k}{K}+n_2+\frac{kq}{K}),
\nonumber\\
&p^2=\frac{1}{\sqrt{2}}(n_1+\frac{k}{K}-n_2-\frac{kq}{K}). 
\end{align} 
The twist boundary condition in the time direction can be implemented by inserting an operator in the Hilbert space. In our case, the desired operator is
\begin{align}
\nonumber \hat{L}(l,k)&=\exp\left[ \frac{2\pi i}{K}\left(lQ_1+qlQ_2\right)\right]\nonumber\\
&=\exp \left[\frac{2\pi i l}{K}\left(n_2+qn_1+\frac{2kq}{K}\right)\right]
\end{align}
where $k,\ l=0,1,2,...,K-1$ and
\begin{align}
Q_1=\frac{1}{\sqrt{2}}(p^1-p^2),\quad Q_2=\frac{1}{\sqrt{2}}(p^1+p^2)
\end{align}
$Q_{1,2}$ satisfies $[\phi_0^1, Q_1]=i$ and $[\phi_0^2, Q_2]=i$ and can generate the translation of $\phi^{1,2}$ in the time direction,

The  partition function for each sector can be calculated by the method in Sec.\ref{edge_cft}. It takes this form
\begin{align}
Z^{k,l}(\tau)&=
\mbox{Tr}_k\left[\hat{L}(l,k)e^{2\pi iP\tau_1-2\pi \tau_2H}\right]
\nonumber \\
&=\frac{1}{|\eta(\tau)|^2}\sum_{n_1,n_2}\exp \left[\frac{2\pi i l}{K}\left(n_2+qn_1+\frac{2kq}{K}\right)\right]
\nonumber \\
&\times 
\exp\left\{-\pi\tau_2\left[\left(n_1+\frac{k}{K}\right)^2+\left(n_2+\frac{qk}{K}\right)^2\right]\right.
\nonumber \\
&\left.+2\pi i\tau_1\left(n_1+\frac{k}{K}\right)\left(n_2+\frac{qk}{K}\right)\right\}.
\label{twist_spt}
\end{align}
Under modular $S$ and $T$ transformations, 
the partition functions are transformed as 
\begin{align}
Z^{k,l}(\tau+1)&=
e^{-\frac{2\pi i qk^2}{K^2}}Z^{k,k+l}(\tau),
\nonumber\\
Z^{k,l}(-1/\tau)&=
e^{\frac{4\pi i qkl}{K^2}}Z^{l,-k}(\tau).
\label{Z_st}
\end{align}

In the next two subsections, we construct the characters for the orbifold CFTs with $\mathbb{Z}_2$ and $\mathbb{Z}_3$ symmetries explicitly. We will also construct the modular invariant partition function with two copies of orbifoldd CFT by following the boson condensation mechanism.

\subsection{$\mathbb{Z}_2$ symmetry}
\label{orb_z2_spt}
There are two inequivalent short-range entangled phases protected 
by $\mathbb{Z}_2$ symmetry.
They are classified by $H^3[\mathbb{Z}_2,U(1)]=\mathbb{Z}_2$ whose cohomological elements are labeled by $q=0,1$. 
It was shown in the Ref.\ \onlinecite{LevinGu12}, after gauging $\mathbb{Z}_2$ symmetry, there are two inequivalent abelian topological phases with different ${\bf K}$ matrices. 
One is the toric code model with ${\bf K}=2\sigma_x$,
and the other is the double semion model with ${\bf K}=2\sigma_z$. 
These two models have the same number of abelian anyons and the same fusion rules. 
However, they have different ${\bf K}$ matrices
and hence different $\mathcal{S}$ and $\mathcal{T}$ matrices. 
In these two models, there are four quasi-particle excitation, including vacuum $1$, bosonic $\mathbb{Z}_2$ charge, $\mathbb{Z}_2$ flux and the flux-charge composite quasi-particle. 
The $\mathbb{Z}_2$ flux is bosonic in the toric code model,
but has conformal spin $h=-1/4$ in the double semion model. 
Furthermore, the flux-charge composite quasi-particle is fermionic in the toric code,
which has conformal spin $h=1/4$ in the double semion model. 

We will have a look at these gauged theories from the point of view of edge theories.  
Using the partition functions calculated in Eq.\eqref{twist_spt}, we can construct the characters for the orbifold CFT. The characters are constructed by applying the $\mathbb{Z}_2$ projection operator on the Hilbert space, which  groups different partition function sectors together as,
\begin{align}
\nonumber \chi_{1}=&\frac{1}{2}\left(Z^{0,0}+Z^{0,1}\right),\\
\nonumber \chi_{2}=&\frac{1}{2}\left(Z^{0,0}-Z^{0,1}\right),\\
\nonumber \chi_{3}=&\frac{1}{2}\left(Z^{1,0}+Z^{1,1}\right),\\
\chi_{4}=&\frac{1}{2}\left(Z^{1,0}-Z^{1,1}\right). 
\end{align}

When $q=0$, the $\mathcal{S}$ and $\mathcal{T}$ matrices are the same as 
those for the toric code model. 
In the toric code model, there are four fundamental excitations, the vacuum sector $\mathbb{I}$, the $\mathbb{Z}_2$ charge $e$, the $\mathbb{Z}_2$ flux $m$ and the flux-charge composite particle $\psi=e\times m$. 
Here we can identify 
$\chi_1=\chi_{\mathbb{I}}$, $\chi_2=\chi_e$, $\chi_3=\chi_m$ 
and $\chi_4=\chi_{\psi}$,
I.e., 
the untwisted sectors correspond to charge excitations, 
while the twisted sectors to flux excitations.
This can be confirmed by calculating the $\mathcal{T}$- and $S$-matrices. 
The total partition function $Z=Z^{0,0}+Z^{0,1}+Z^{1,0}+Z^{1,1}$ is modular invariant and this implies that the SPT phase (with $q=0$) before gauging is trivial.

On the other hand, if $q=1$, the orbifold CFT corresponds to the edge of double semion model. 
The double semion model has four quasi-particle excitations $\mathbb{I}, \bar{s}, s, \bar{s}$, where $\mathbb{I}$ is the vacuum sector, $s\bar{s}$ is the $\mathbb{Z}_2$ boson, $\bar{s}$ is the $\mathbb{Z}_2$ flux and $s$ is the flux-charge composite particle. 
The characters are
\begin{align}
\nonumber \chi_{\mathbb{I}}&=\chi_1=\frac{|\Theta_{0,2}|^2}{|\eta(\tau)|^2},
\\
\nonumber \chi_{s\bar{s}}&=\chi_2=\frac{|\Theta_{1,2}|^2}{|\eta(\tau)|^2},
\\
\nonumber \chi_{\bar{s}}&=\chi_3=\frac{\Theta_{0,2}\overline{\Theta}_{1,2}}{|\eta(\tau)|^2},
\\
\chi_{s}&=\chi_4=\frac{\Theta_{1,2}\overline{\Theta}_{0,2}}{|\eta(\tau)|^2},
\end{align}
where $\Theta_{\lambda, K}$ is defined in Eq.\eqref{theta_f} and $\eta$ is defined in Eq.\eqref{eta_func}. Since $\Theta_{\lambda,K}/\eta$ is the character for the chiral $U(1)_{K/2}$ CFT, the above expressions are the characters for
the $U(1)_1\times\overline{U(1)}_1$ CFT which 
is the edge theory of the double semion model.\cite{LevinWen05}  
The total partition function for the edge 
of the double semion model is constructed by 
taking the linear combination of $Z^{k,l}$ with $k,l=0,1$.
By using Eq.\eqref{Z_st}, one can readily check that $Z=\sum_{k,l}\varepsilon(k,l)Z^{k,l}$ cannot be modular invariant, where $\varepsilon(k,l)$ is an arbitrary phase.\cite{Sule_2013}
This leads us to conclude that the SPT phase before gauging is non-trivial, 
and its gapless edge state is protected by $\mathbb{Z}_2$ symmetry. 
We need at least two copies of the SPT phases 
to gap out the edge without breaking the symmetry.
After gauging $\mathbb{Z}_2$ symmetry, 
the two copies of the non-trivial SPT model become two copies of the double semion model. 
The edge theory partition function can be constructed and checked that it
can be made modular invariant.

Alternatively, this can be understood through the boson condensation mechanism in 
the two copies of the double semion model as follows. 
This model has 16 quasi-particle excitations, 
where the quasi-particle 
$s_1\bar{s}_1s_2\bar{s}_2$ is a boson.
The condensation of this boson identifies $\psi=s_1s_2\equiv \bar{s}_1\bar{s}_2$, $e=s_1\bar{s}_1\equiv s_2\bar{s}_2$ and $m=\bar{s}_1s_2\equiv s_1\bar{s}_2$. 
All the other excitations which have a non-trivial braiding phase 
with $s_1\bar{s}_1s_2\bar{s}_2$ are confined. 
This topological phase after condensation has four quasi-particle excitations 
and their braiding statistics is equivalent to that of the toric code model. 
The edge theory characters can be constructed from the original characters as,
\begin{align}
&\widetilde{\chi}_I=\chi_I\chi_I+\chi_{s\bar{s}}\chi_{s\bar{s}},\nonumber\\
&\widetilde{\chi}_e=\chi_{s\bar{s}}\chi_I+\chi_{s\bar{s}}\chi_I,\nonumber\\
&\widetilde{\chi}_m=\chi_{\bar{s}}\chi_s+\chi_s\chi_{\bar{s}},\nonumber\\
&\widetilde{\chi}_{\psi}=\chi_{s}\chi_s+\chi_{\bar{s}}\chi_{\bar{s}}.
\end{align} 
It is easy to confirm that these four characters have the same $\mathcal{S}$ and $\mathcal{T}$ matrices as the toric code model, 
and also 
the total partition function  $Z=(Z^{0,0})^2+(Z^{0,1})^2+(Z^{1,0})^2-(Z^{1,1})^2$ 
is invariant under the modular transformation.

\subsection{$\mathbb{Z}_3$ symmetry}
\label{orb_z3_spt}
There are three inequivalent SPT phases with $\mathbb{Z}_3$ symmetry which are classified by $H^3[\mathbb{Z}_3,U(1)]=\mathbb{Z}_3$ whose cohomological elements are labeled by $q=0,\ 1$ and 3. After gauging $\mathbb{Z}_3$ symmetry, there are three inequivalent abelian topological phases with different $\textbf{K}$ matrices. 
Here we will  gauge the $\mathbb{Z}_3$ symmetry for both the non-trivial SPT phase and the trivial SPT phase.

\subsubsection{gauging non-trivial $\mathbb{Z}_3$ SPT} 
Similar to the edge of the non-trivial $\mathbb{Z}_2$ SPT phase, the non-trivial $\mathbb{Z}_3$ SPT phase can also be described by a Luttinger liquid with $\textbf{K}=\sigma_x$. This edge CFT is protected by $\mathbb{Z}_3$ symmetry, which is 
\begin{align}
\phi^1\to\phi^1+2\pi k/3,\quad \phi^2\to\phi^2+2\pi kq/3
\label{symm_z3}
\end{align}
where $k,q=0,1,2$. 

\paragraph{$q=1$}

Here we first consider $q=1$ and the partition function is
\begin{align}
Z^{k,l}&=|\eta(\tau)|^{-2}\sum_{n,m}\exp\left\{\frac{2\pi il}{K}(m+n+\frac{2k}{K})\right.
\nonumber\\
&
\quad 
\left.-\pi\tau_2\left[(n+\frac{k}{K})^2+(m+\frac{k}{K})^2\right]
\right.
\nonumber \\
&
\quad
\left.
+2\pi i\tau_1(n+\frac{k}{K})(m+\frac{k}{K})\right\}
\end{align}
where $k,l=0,1,2$. 
$k$ and $l$ label the twist in the spatial and time directions,  respectively.
These partition functions can be used to construct the characters for the orbifold CFT, which is the effective edge state of topological phase after gauging. These characters are listed in the Table \eqref{z3_spt}, where $z$ represents the $\mathbb{Z}_3$ charge and $\bar{z}$ is its anti-particle. $\rho_0$ is the $\mathbb{Z}_3$ flux, $\rho_1=\rho_0\times z$ and $\rho_2=\rho_0\times\bar{z}$ are the flux-charge composite particles. $\bar{\rho}_j$ is the anti-particle of $\rho_j$ ($j=0,1,2$). 
By checking the $\mathcal{S}$ and $\mathcal{T}$ matrices for the orbifold CFT, we find that this actually is the effective edge theory of an abelian topological phase with the  {K}-matrix
\begin{align}
{\bf K}=\begin{pmatrix} 4&5\\ 5&4\end{pmatrix}. 
\label{Z_3_1}
\end{align}
All the quasi-particle excitations can be written as $\alpha{\bf e}_1$, where ${\bf e}_1=(1,0)$ is the fundamental anyon lattice vector for the above {K}-matrix. The anyon lattice has fusion group $\mathbb{Z}_9$ where three gauge fluxes fuse into a charge conjugate $\bar{z}$ rather than the vacuum. This is different from fusion rule in the ordinary $D(\mathbb{Z}_3)$ quantum double model, which has fusion group $\mathbb{Z}_3\times\mathbb{Z}_3$ and is the topological phase after gauging the trivial $\mathbb{Z}_3$ SPT phase.\cite{BaisDrielPropitius92, Kitaev97, Bais-2007, Propitius-1995, Preskilllecturenotes, Freedman-2004, Mochon04}  For the non-trivial $\mathbb{Z}_3$ SPT phase, the edge partition function after orbifolding is $Z=\sum_{k,l}\varepsilon(k,l) Z^{k,l}$ where $k,\ l=0, 1, 2$ and $\varepsilon(k,l)$ is an arbitray phase. For this partition function, according to Eq.\eqref{Z_st},  under $\mathcal{T}$ transformation, there is a different phase for $Z^{k,l}$ with different $k$ and $Z$ cannot be modular invariant. Therefore the SPT phase before gauging is non-trivial.

\begin{table}[t]
\centering
\begin{ruledtabular}
\begin{tabular}{lcc}
character $\chi$ & $\textbf{l}_{\chi}$ & $h_\chi$ \\\hline
$\chi_{\mathbb{I}}=Z^{0,0}+Z^{0,1}+Z^{0,2}$ & $0$ & $0$\\
$\chi_{z}=Z^{0,0}+\omega^2 Z^{0,1}+\omega Z^{0,2}$ & $6{\bf e}_1$ & $0$\\
$\chi_{\bar{z}}=Z^{0,0}+\omega Z^{0,1}+\omega^2 Z^{0,2}$ & $3{\bf e}_1$ & $0$\\
$\chi_{\rho_0}=Z^{1,0}+e^{-\frac{4\pi i}{9}}Z^{1,1}+e^{-\frac{8\pi i}{9}}Z^{1,2}$ & $7{\bf e}_1$ & ${1}/{9}$\\
$\chi_{\rho_1}=Z^{1,0}+\omega^2 e^{-\frac{4\pi i}{9}}Z^{1,1}+\omega e^{-\frac{8\pi i}{9}}Z^{1,2}$ & $4{\bf e}_1$ & ${4}/{9}$\\
$\chi_{\rho_2}=Z^{1,0}+\omega e^{-\frac{4\pi i}{9}}Z^{1,1}+\omega^2 e^{-\frac{8\pi i}{9}}Z^{1,2}$ & ${\bf e}_1$ & ${7}/{9}$\\
$\chi_{\bar{\rho}_0}=Z^{2,0}+\omega e^{-\frac{2\pi i}{9}}Z^{2,1}+\omega^2 e^{-\frac{4\pi i}{9}}Z^{2,2}$ & $2{\bf e}_1$ & ${1}/{9}$\\
$\chi_{\bar{\rho}_1}=Z^{2,0}+\omega^2 e^{-\frac{2\pi i}{9}}Z^{2,1}+\omega e^{-\frac{4\pi i}{9}}Z^{2,2}$ & $5{\bf e}_1$ & ${4}/{9}$\\
$\chi_{\bar{\rho}_2}=Z^{2,0}+e^{-\frac{2\pi i}{9}}Z^{2,1}+e^{-\frac{4\pi i}{9}}Z^{2,2}$ & $8{\bf e}_1$ & ${7}/{9}$\\
\end{tabular}
\end{ruledtabular}
\caption{The characters for 
orbifolding the $\mathbb{Z}_3$ symmetry in the $\mathbb{Z}_3$ non-trivial SPT phase with $q=1$. 
${\bf l}_{\chi}$ is the corresponding anyon lattice vector with ${\bf e}_1=(1,0)$, 
$h$ is the conformal dimension (mod $\mathbb{Z}$), and $\omega=e^{2\pi i/3}$.
}\label{z3_spt}
\end{table}

\paragraph{$q=2$}

For Eq.\ \eqref{symm_z3}, when $q=2$, the SPT phase is also non-trivial. We can use similar method to construct the characters for the $\mathbb{Z}_3$ orbifold CFT.  By calculating the modular $\mathcal{S}$ and $\mathcal{T}$ matrices on the boundary, we can show that the topological phase in the bulk has the  {K}-matrix equal to (conjugate to $q=1$ case)
\begin{align}
{\bf K}=\begin{pmatrix} -4&5\\ 5&-4\end{pmatrix}
\label{Z_3_2}
\end{align}

\paragraph{$q=0$}

When $q=0$, the SPT phase is trivial and the partition function for the $\mathbb{Z}_3$ orbifold CFT is modular invariant. The $\mathcal{S}$ and $\mathcal{T}$ matrices for the $\mathbb{Z}_3$ orbifold CFT indicate that the topological phase after gauging is $D(\mathbb{Z}_3)$ quantum double model with ${\bf K}=3\sigma_x$. From the above discussion, we show that $q=0,1,2$ correspond to three different SPT phases and the gauged topological phase correspond to three different abelian topological phases with different {\bf K}-matrices. This result is consistent with the result from group cohomology classification that  $H^3(\mathbb{Z}_3,U(1))=\mathbb{Z}_3$.

The collection of the three distinct topological phases has a $\mathbb{Z}_3$ tensor product structure. It can be understood through boson condensation.\cite{BaisSlingerlandCondensation,Kong14} Assume we put together two copies of topological phase with {K}-matrix defined in Eq.\eqref{Z_3_1} by tensor product, the composite excitation is therefore denoted as  $\alpha_i^1\alpha_j^2$, where $\alpha$ is the excitation in each layer. After condensing the mutually local boson pairs $z^1z^2$ and $\bar{z}^1\bar{z}^2$, since $z^1$, $\bar{z}^2$ and $\bar{z}^1z^2$ are differed from each other by the boson $z^1z^2$ or $\bar{z}^1\bar{z}^2$, the pure boson $z^1$ is identified with $\bar{z}^2$ and $\bar{z}^1z^2$. Similarly, we also have $\bar{z}^1\sim z^2\sim z^1\bar{z}^2$. The $\mathbb{Z}_3$ flux can pair together to form the excitation in the condensed phase and the complete result is listed in Table.\eqref{deconfined_ex}. 
The new topological phase actually is the other gauged non-trivial $\mathbb{Z}_3$ SPT ($q=2$) with the {K}-matrix in Eq.\ \eqref{Z_3_2}. There are in total nine deconfined excitation and all the other excitations having non-zero braiding phase with $z^1z^2$ or $\bar{z}^1\bar{z}^2$ are confined. 

We can add another layer with the {K}-matrix defined 
in Eq.\ \eqref{Z_3_1} to the double-layer system and further condense the boson $z^1z^3$ and $\bar{z}^1\bar{z}^{3}$, the final topological phase has ${\bf K}=3\sigma_x$ which is the topological phase after gauging the trivial $\mathbb{Z}_3$ SPT phase.

The above boson condensation in the bulk topological phase can also be understood by looking at the edge CFT. The condensation process is the reversed process of orbifolding and we can use the characters listed in Table \eqref{z3_spt} to form the new characters for the {K}-matrix in Eq.\ \eqref{Z_3_2} and $3\sigma_x$. Since a similar calculation has already been done in Sec.\ \ref{orb_z2_spt}, we will not construct the characters explicitly here. 

\begin{table}[t]
\centering
\begin{ruledtabular}
\begin{tabular}{ l l c }
  deconfined excitation & lattice vector & $h$  \\\hline
  $\mathbb{I}\sim z^1z^2\sim \bar{z}^1\bar{z}^2$ & $\mathbb{I}=0$ & 0  \\
  $z^1\sim \bar{z}^2\sim \bar{z}^1 z^2$ & $z=3{\bf e}_1$ & 0  \\
  $\bar{z}^1\sim z^2\sim z^1\bar{z}^2$ & $\bar{z}=6{\bf e}_1$ & 0\\
  $\rho_0^1\bar{\rho}_0^2\sim\rho_1^2\bar{\rho}_2^2\sim\rho_2^1\bar{\rho}_1^2$ & $\rho_0={\bf e}_1$& ${2}/{9}$\\
  $\rho_2^1\bar{\rho}_2^2\sim\rho_0^2\bar{\rho}_1^2\sim\rho_1^1\bar{\rho}_0^2$ & $\rho_1=4{\bf e}_1$ & ${5}/{9}$\\
  $\rho_1^1\bar{\rho}_1^2\sim\rho_2^2\bar{\rho}_0^2\sim\rho_0^1\bar{\rho}_2^2$ & $\rho_2=7{\bf e}_1$& ${8}/{9}$\\
  $\bar{\rho}_0^1{\rho}_0^2\sim\bar{\rho}_1^2{\rho}_2^2\sim\bar{\rho}_2^1{\rho}_1^2$ & $\bar{\rho}_0=8{\bf e}_1$ & ${2}/{9}$\\
  $\bar{\rho}_2^1{\rho}_2^2\sim\bar{\rho}_0^2{\rho}_1^2\sim\bar{\rho}_1^1{\rho}_0^2$ & $\bar{\rho}_1=5{\bf e}_1$ & ${5}/{9}$\\
  $\bar{\rho}_1^1{\rho}_1^2\sim\bar{\rho}_2^2{\rho}_0^2\sim\bar{\rho}_0^1{\rho}_2^2$ & $\bar{\rho}_2=2{\bf e}_1$& ${8}/{9}$\\
\end{tabular}
\end{ruledtabular}
\caption{The first column is the deconfined excitation table after boson condensation, the second column is the corresponding anyon lattice vector in the topological phase with the {K}-matrix in Eq.\eqref{Z_3_2} and $h$ is the conformal dimension (mod $\mathbb{Z}$).
}\label{deconfined_ex}
\end{table}

\subsubsection{gauging trivial $\mathbb{Z}_3$ SPT}

In this subsection, we discuss another $\mathbb{Z}_3$ SPT phase. 
We will show that this SPT phase is trivial by studying the edge $\mathbb{Z}_3$ orbifold CFT. 
This model is described by the following {K}-matrix: 
\begin{align}
\textbf{K}=k\begin{pmatrix}0&0&0&-1\\0&0&1&0\\0&1&0&0\\-1&0&0&0\end{pmatrix}
\label{ksptZ_3}
\end{align}
This model is discussed in Ref.\ \onlinecite{BombinMartin06, TeoRoyXiao13long} and can be realized as an exactly solvable spin (rotor) model on the honeycomb lattice. It has $k^4$ abelian anyon excitations and can be labeled by ${\bf a}=({\bf a}_{\bullet},
{\bf a}_{\circ})$. Each component ${\bf a}_{\bullet/\circ}$ lives on a two dimensional triangular $\mathbb{Z}_k$ lattice and has threefold rotation symmetry. In total, the anyon excitation has $S_3=\mathbb{Z}_3\ltimes \mathbb{Z}_2$, where the extra $\mathbb{Z}_2$ symmetry is coming from the twofold interchange $\bullet\leftrightarrow\circ$ symmetry. Here we consider a trivial example with $k=1$, so that it does not have topological order and can only be a SPT phase. We will gauge the $\mathbb{Z}_3$ symmetry operator and study the orbifold CFT on the boundary. The $\mathbb{Z}_3$ symmetry operator is defined to be 
\begin{align}
{\Lambda_3}=\begin{pmatrix}0&-1&0&0\\1&-1&0&0\\0&0&0&-1\\0&0&1&-1\end{pmatrix}
\end{align}
which satifies $\Lambda_3^3=\mathbb{I}$ and $\Lambda_3^T\textbf{K}\Lambda_3=\textbf{K}$. The eigenvalues of $\Lambda_3$ are $\omega=e^{2\pi i/3}$ and $\omega^2$.

Similar to the method used in the last section, we can define the new bosonic field $\vec{\varphi}$ to diagonalize the 
$\textrm{K}$-matrix. 
Under $\mathbb{Z}_3$ symmetry operation, $\varphi^1+i\varphi^2\to\omega^n(\varphi^1+i\varphi^2)$ and $\varphi^3+i\varphi^4\to\omega^n(\varphi^3+i\varphi^4)$, where $n=1,2$. By considering twist boundary conditions in spatial and time direction, we will get different sectors of partition function $Z^{\mu\nu}$, where $\mu$ and $\nu$ represent the twist boundary condition in time and space direction respectively. $\mu,\nu=0,\frac{1}{3},\frac{2}{3}$. 
The partition function for each twist sector equals to \cite{bigyellowbook}
\begin{align}
Z^{\mu,\nu}=\left|\frac{\eta(\tau)}{\theta_{\beta}^{\alpha}(\tau)}\right|^2
\end{align}
where $\alpha=\frac{1}{2}-\mu$ and $\beta=\frac{1}{2}+\nu$.
The $\theta^{\alpha}_{\beta}(\tau)$ function is 
defined as 
\begin{align}
\theta^{\alpha}_{\beta}(\tau)&=\theta\begin{bmatrix} \alpha \\ \beta \end{bmatrix}(\tau,0)=\sum_{n\in \mathbb{Z}}q^{\frac{1}{2}(n+\alpha)^2}e^{2\pi i(n+\alpha)\beta}
\nonumber \\
&=\eta(\tau)e^{2\pi i\alpha\beta}q^{\frac{\alpha^2}{2}-\frac{1}{24}}
\nonumber \\
&\quad 
\times 
\prod_{n=1}^{\infty}(1+q^{n+\alpha-\frac{1}{2}}e^{2\pi i\beta})(1+q^{n-\alpha-\frac{1}{2}}e^{-2\pi i\beta}).
\label{theta}
\end{align}
Under the $S$ and $T$ transformations, it transforms as
\begin{align}
&\theta_{\beta}^{\alpha}(\tau+1)=
e^{-\pi i\alpha(\alpha-1)}\theta_{\alpha+\beta-\frac{1}{2}}^{\alpha}(\tau),
\nonumber \\
&\theta_{\beta}^{\alpha}(-{1}/{\tau})=\sqrt{-i\tau}e^{2\pi i\alpha\beta}\theta_{-\alpha}^{\beta}(\tau). 
\label{thetast}
\end{align}

Thus we can get the modular transformation for the partition function. 
Since $\mu\in [0,1)$ and $\nu\in [0,1)$, we have 
\begin{align}
&Z^{\mu,\nu}(\tau+1)=Z^{\mu,[\nu-\mu]}(\tau),
\nonumber \\
&Z^{\mu,\nu}(-{1}/{\tau})=Z^{[1-\nu],\mu}(\tau),
\end{align}
where $[\nu-\mu]$ is $(\nu-\mu)$ modulo $1$ and $[1-\nu]$ is $(1-\nu)$ modulo $1$. 

According to the definition of character in Eq.\ \eqref{char_def}, we group the partition functions into characters and list the result in the following align,
\begin{align}
\chi_{\mathbb{I}}&=\frac{1}{3}(Z_0+Z^{0,\frac{1}{3}}+Z^{0,\frac{2}{3}}),
\nonumber \\
\chi_z&=\frac{1}{3}(Z_0+\omega^{-1}Z^{0,\frac{1}{3}}+\omega Z^{0,\frac{2}{3}}),
\nonumber \\
\chi_{\bar{z}}&=\frac{1}{3}(Z_0+\omega Z^{0,\frac{1}{3}}+\omega^{-1} Z^{0,\frac{2}{3}}),
\nonumber \\
\chi_{\rho_0}&=\frac{1}{3}(Z^{\frac{2}{3},0}+Z^{\frac{2}{3},\frac{1}{3}}+Z^{\frac{2}{3},\frac{2}{3}}),
\nonumber \\
\chi_{\rho_1}&=\frac{1}{3}(Z^{\frac{2}{3},0}+\omega^{-1}Z^{\frac{2}{3},\frac{1}{3}}+\omega\ Z^{\frac{2}{3},\frac{2}{3}}),
\nonumber \\
\chi_{\rho_2}&=\frac{1}{3}(Z^{\frac{2}{3},0}+\omega Z^{\frac{2}{3},\frac{1}{3}}+\omega^{-1} Z^{\frac{2}{3},\frac{2}{3}}),
\nonumber \\
\chi_{\bar{\rho}_0}&=\frac{1}{3}(Z^{\frac{1}{3},0}+Z^{\frac{1}{3},\frac{1}{3}}+Z^{\frac{1}{3},\frac{2}{3}}),\nonumber \\
\chi_{\bar{\rho}_1}&=\frac{1}{3}(Z^{\frac{1}{3},0}+\omega Z^{\frac{1}{3},\frac{1}{3}}+\omega^{-1}Z^{\frac{1}{3},\frac{2}{3}}),
\nonumber \\
\chi_{\bar{\rho}_2}&=\frac{1}{3}(Z^{\frac{1}{3},0}+\omega^{-1}Z^{\frac{1}{3},\frac{1}{3}}+\omega Z^{\frac{1}{3},\frac{2}{3}}), 
\end{align}
where $z$ is the $\mathbb{Z}_3$ boson and $\bar{z}$ is its anti-particle. $\rho_0$ is the $\mathbb{Z}_3$ flux and $\rho_1=\rho_0\times z$ and $\rho_2=\rho_0\times \bar{z}$ are the flux-particle composite particles. $\bar{\rho}_j$ are the anti-particle of $\rho_j$ with $j=0,1,2$.

By further checking the $\mathcal{S}$ and $\mathcal{T}$ matrices, we notice that that are the same as that for the $D(\mathbb{Z}_3)$ quantum double model with ${\bf K}=3$. The correspondence between $\mathbb{Z}_3$ orbifold CFT and the edge CFT for $D(\mathbb{Z}_3)$ quantum double model is shown in Table \ref{SPTZ_3}. The conformal dimensions $h$ for these characters are also listed in this table.
\begin{table}[htbp]
\centering
\begin{ruledtabular}
\begin{tabular}{lcc}
Orbifolding $\mathbb{Z}_3$  & $D(\mathbb{Z}_3)$ & $h$  \\\hline
$\chi_{\mathbb{I}}$ & $(0,0)$ & $0$\\
$\chi_z$ & $(0,1)$ & $0$\\
$\chi_{\bar{z}}$ &$(0,2)$ & $0$\\
$\chi_{\rho_0}$ & $(1,0)$ & $0$\\
$\chi_{\rho_1}$ & $(1,1)$ & ${1}/{3}$\\
$\chi_{\rho_2}$ & $(1,2)$ & ${2}/{3}$\\
$\chi_{\overline{\rho}_0}$ & $(2,0)$ & $0$\\
$\chi_{\overline{\rho}_1}$ & $(2,2)$ & ${1}/{3}$\\
$\chi_{\overline{\rho}_2}$ & $(2,1)$ & ${2}/{3}$\\
\end{tabular}
\end{ruledtabular}
\caption{The characters after gauging $\mathbb{Z}_3$ symmetry and the corresponding anyon lattice vector for $D(\mathbb{Z}_3)$ quantum double model. $h$ is the conformal dimension (mod $\mathbb{Z}$).
}\label{SPTZ_3}
\end{table}



Actually, the $\mathrm{K}$-matrix defined in Eq.\eqref{ksptZ_3} with $\Lambda_3$ symmetry is a trivial SPT phase. It is similar to the trivial SPT phase with $q=0$ defined in the previous section. The total partition function $Z_{tot}=\sum_{\mu\nu}Z^{\mu\nu}=\chi_I+\chi_{\rho_0}+\chi_{\overline{\rho}_0}$  is modular invariant.

The gapless edge CFT of this trivial phase can be gapped out be by a potential term. The potential term we choose takes this form,
\begin{align}
W=-\cos(\vec{l}_1^T\vec{\Phi})-\cos(\vec{l}_2^T\vec{\Phi})-\cos(\vec{l}_3^T\vec{\Phi})
\end{align}
where $\vec{l}_1^T=(1,1,1,1)$, $\vec{l}_2^T=(-1,0,-1,0)$ and $\vec{l}_3^T=(0,-1,0,-1)$. Since $\Lambda_3\vec{l}_1=\vec{l}_2$, $\Lambda_3\vec{l}_2=\vec{l}_3$ and $\Lambda_3\vec{l}_3=\vec{l}_1$, the potential W is invariant under $\Lambda_3$ symmetry. Moreover,
\begin{align}
[\vec{l}_i^T\partial_x\vec{\Phi}(x),\vec{l}_j^T\vec{\Phi}(y)]=2\pi i (\vec{l}_i^T\textbf{K}^{-1}\vec{l}_j)\delta(x-y)=0.
\end{align}
$\cos(\vec{l}_1^T\vec{\Phi})$, $\cos(\vec{l}_2^T\vec{\Phi})$ and $\cos(\vec{l}_3^T\vec{\Phi})$ can be simultaneously gapped without breaking the symmetry.

\section{Topologically ordered phases}
\label{SET_orbifold}

In this section, we discuss gauging symmetries in topologically ordered phases,
from the point of view of boundary (conformal) field theories. 
We will consider three different abelian topological order described by 
the K-matrix abelian Chern-Simons theories, 
\eqref{K_Zn} (the $D(\mathbb{Z}_N)$ quantum double model),
\eqref{K_su3} ($SU(3)_1$),
and 
\eqref{K_so8} ($SO(8)_1$).
By subsequently taking their orbifold theories along the system boundary, we will obtain the topological content of the $(2+1)$d non-abelian twist liquid after gauging their symmetries.

\subsection{Orbifolding $\mathbb{Z}_2$ symmetry in the $D(\mathbb{Z}_N)$ quantum double model}

The $D(\mathbb{Z}_N)$ quantum double model describes an abelian topological order  
realized in the deconfined phase of 
the $2+1$-dimensional $\mathbb{Z}_N$ gauge theory. 
Equivalently, this topological order can also 
be described by the abelian Chern-Simons theory 
in Eq.\ \eqref{chern_simons} with ${\bf K}=N\sigma_x$,
\begin{align}
\textbf{K}=
\begin{pmatrix} 0 & N \\ N & 0 \end{pmatrix}.
\label{K_Zn}
\end{align}
There are two fundamental quasi-particle excitations in this model:
$e$-particle corresponding to the quasiparticle lattice vector $\textbf{e}^T=(1,0)$,
and 
$m$-particle corresponding to $\textbf{m}^T=(0,1)$. 
All the quasi-particles can be written as 
a linear combination of $e$- and $m$-particles, 
$a{\bf e}+b{\bf m}$, 
where $0\leq a<N$ and $0\leq b<N$. 
$e$- and $m$-particles are self-bosons, 
and obey non-trivial mutual braiding statistics 
with the $e^{2\pi i/N}$ braiding phase 

The $D(\mathbb{Z}_N)$ quantum double model has a global $e$-$m$ duality which 
exchanges $e$- and $m$-particles. 
This duality symmetry is an anyonic symmetry 
and leaves $\mathcal{S}$ and $\mathcal{T}$ matrices invariant. 
By gauging, the $\mathbb{Z}_2$ global duality symmetry can be promoted to a local symmetry.\cite{BarkeshliBondersonChengWang14} 
The $\mathbb{Z}_2$ charge $c$ is a boson and satisfies $c\times c=1$. 
The $\mathbb{Z}_2$ invariant state $\psi_r=e^rm^r$ will split into two states $\psi_{r^\pm}$, 
which differ by a $\mathbb{Z}_2$ charge, 
${\psi}_{r^-}=c\times \psi_{r^+}$. 
On the other hand, 
the $\mathbb{Z}_2$ non-invariant states $e^am^b$ will be able to rotate into each other under the $\mathbb{Z}_N$ symmetry, and will therefore be needed to be grouped together to form a $\mathbb{Z}_2$ invariant superselection sector denoted by
with quantum dimension 2. 
The $\mathbb{Z}_2$ flux $\sigma$
is a non-abelian quasi-particle and can carry $N$ different species labels. This is because $\sigma_{\lambda+a} = \sigma_\lambda \times e^a = \sigma_\lambda \times m^a$, for $a=0,1,\ldots,N-1$  and $\lambda$ represents the species labels.
When $N$ is even, $\lambda$ takes an integer value, when $N$ is odd, $\lambda$ takes a half-integer value.\cite{TeoRoyXiao13long} In addition, 
The $\mathbb{Z}_2$ flux can combine with $c$ to form a flux-charge composite particle.
In total, after gauging, there are $N(N+7)/2$ quasi-particles (Table \ref{toric_even}, \ref{toric_odd}).
To fully understand the fusion and braiding statistics for these anyons, 
below we will study the edge CFT before and after orbifolding the $\mathbb{Z}_2$ duality symmetry.


\subsubsection{Characters for the $D(\mathbb{Z}_N)$ quantum double model}

The edge CFT can be described by Eq.\ \eqref{lutt} 
with two bosonic fields $\phi_1$ and $\phi_2$ and ${\bf K}=N\sigma_x$. 
Following the general method discussed in Sec.\ \ref{edge_cft}, 
the {K}-matrix can be diagonalized as $\textbf{K}=U^{T}\eta U$, 
where 
$U=\sqrt{\frac{N}{2}}\begin{pmatrix} 1 & 1 \\ 1 & -1 \end{pmatrix}$ and $\eta=\begin{pmatrix} 1 & 0 \\ 0 & -1 \end{pmatrix}$. 
After diagonalizing the {K}-matrix, 
there are two new bosonic fields, $\varphi_1$ and $\varphi_2$, 
with the zero mode momentum 
$p^1=\sqrt{\frac{1}{2N}}(Ns+a+Nt+b)$ and $p^2=\sqrt{\frac{1}{2N}}(Ns+a-Nt-b)$, 
where $\vec{\Lambda}^T=(s,t)\in\mathbb{Z}^2,\quad 0\leq a<N$ and $0\leq b<N$.
The characters for the $D(\mathbb{Z}_N)$ quantum double model are labeled by $a$ and $b$,
\begin{align}
X_{a,b}(\tau)=\frac{1}{|\eta(\tau)|^2}\sum_{s,t}q^{\frac{1}{4N}(Ns+a+Nt+b)^2}\bar{q}^{\frac{1}{4N}(Ns+a-Nt-b)^2}. 
\label{char_toric}
\end{align}
There are $N^2$ characters in total. 
Under $T$ and $S$ transformations, 
these characters are transformed as 
\begin{align}
X_{a,b}(\tau+1)&=e^{2\pi i\frac{ab}{N}}X_{a,b}(\tau),
\nonumber \\
X_{a,b}(-1/\tau)&=\frac{1}{N}\sum_{a^{\prime},b^{\prime}}X_{a^{\prime},b^{\prime}}e^{-2\pi i\frac{a^{\prime}b+b^{\prime}a}{N}}. 
\end{align}

\subsubsection{Orbifolding $\mathbb{Z}_2$ symmetry when $N$ is even}

In the language of the $2+1$-dimensional Chern-Simons field theory, 
the $\mathbb{Z}_2$ duality symmetry is represented by the Pauli matrix $\sigma_x$.
The $\mathrm{K}$-matrix is invariant under the duality symmetry as $\textbf{K}=\sigma_x^T\textbf{K}\sigma^{\ }_x$. 
Thus the duality symmetry is an anyonic symmetry and leaves $\mathcal{S}$ and $\mathcal{T}$ matrices invariant. 
Correspondingly, in terms of the edge CFT, 
the bosonic fields are transformed 
under the two-fold symmetry as $\phi_1\leftrightarrow\phi_2$. 
In the $\vec{\varphi}$ basis, the transformation law reads
$\varphi_1\to\varphi_1$ and $\varphi_2\to-\varphi_2$. 
This means that, when orbifolding $\mathbb{Z}_2$ symmetry, 
we need to consider the twisted boundary condition for $\varphi_2$ field 
both in time and spatial directions.

For the untwisted boundary condition, the partition function is the same as  
the $D(\mathbb{Z}_N)$ model.
As for the twist boundary condition in time direction, 
\begin{align}
\varphi_2(x,t+2\pi)=-\varphi_2(x,t), 
\end{align}
it is equivalent to insert the $\mathbb{Z}_2$ symmetry operator $\sigma$ in the partition function
to remove all the states which are not invariant under $\sigma$. 
Focusing on the zero-mode part, 
this operator will remove all the zero momentum modes which are not invariant under the $\sigma$ symmetry operator. 
Only the states which satisfy $\sigma|\textbf{K}\vec{\Lambda}+\vec{\lambda}\rangle=|\textbf{K}\vec{\Lambda}+\vec{\lambda}\rangle$ will remain, 
which requires that $s=t$ and $a=b$ and restricts the summation in Eq.\eqref{char_toric}. 
The original zero mode partition function therefore becomes $\Theta_{2r,2N}(\tau)$ 
which is the zero mode part for the chiral $U(1)_N$ CFT,
Eq.\ \eqref{theta_f}.
As for the oscillator part in the mode expansion of $\varphi_2$, 
under $\mathbb{Z}_2$ symmetry operator, $\varphi_2\to -\varphi_2$, this contributes a term $\sqrt{\eta(\bar{\tau})/\theta_2(\bar{\tau})}$, which is the partition function for the chiral free boson with $\mathbb{Z}_2$ twist in the time direction.\cite{bigyellowbook} 
Combined with the oscillator part from $\varphi_1$,  
the total partition function with twist in time direction therefore is given by
\begin{align}
\label{Z_01}
\mathcal{Z}_r^{0,\frac{1}{2}}=\frac{1}{\eta(\tau)}\Theta_{2r,2N}(\tau)\sqrt{\frac{\eta(\bar{\tau})}{\theta_2(\bar{\tau})}}
\end{align}
where $1/\eta(\tau)$ is coming from the oscillator of $\varphi_1$ field and we have $0\leq r<N$.

As for the twist boundary condition in space direction, 
\begin{align}
\varphi_2(x+2\pi,t)=-\varphi_2(x,t), 
\end{align}
By simply performing modular $S$ transformation on $\Theta_{2r,2N}(\tau)$ of $\mathcal{Z}_r^{0,\frac{1}{2}}$ in Eq.\eqref{Z_01}, we can obtain the zero mode partition function $\Theta_{l,2N}(\tau)+\Theta_{l+N,2N(\tau)}$.  An alternative way to get this result is by calculating zero mode partition function for only $\varphi^1$ field with periodic boundary condition. In addition, the oscillator part for the $\varphi_2$ field will give rise to $\sqrt{\eta(\bar{\tau})/\theta_4(\bar{\tau})}$ and the total partition function is given by
\begin{align}
\mathcal{Z}_l^{\frac{1}{2},0}=\frac{1}{\eta(\tau)}
\left[
\Theta_{l,2N}(\tau)+\Theta_{l+N,2N}(\tau)
\right] 
\sqrt{\frac{\eta(\bar{\tau})}{\theta_4(\bar{\tau})}}, 
\end{align}
where $0\leq l<N$. 

Finally, as for the boundary condition twisted in both time and space directions, 
the partition function can be obtained by applying $T$ transformation on $\mathcal{Z}_l^{\frac{1}{2},0}$.
The resulting partition function is given by 
\begin{align}
\mathcal{Z}_l^{\frac{1}{2},\frac{1}{2}}
&=\frac{1}{\eta(\tau)}
\left[
\Theta_{l,2N}(\tau)+(-1)^{\frac{N}{2}+l}\Theta_{l+N,2N}(\tau)
\right]
\sqrt{\frac{\eta(\bar{\tau})}{\theta_3(\bar{\tau})}}
\end{align}
with $0\leq l<N$. 
The $\theta_{2,3,4}$ functions used in the above partition functions is actually a special case of $\theta^{\alpha}_{\beta}(\tau)$ function defined in Eq.\eqref{theta}, and are related as
\begin{align}
\theta_2=\theta^{1/2}_0,\quad \theta_4=\theta^{0}_{1/2},\quad \theta_3=\theta^0_0.
\end{align}
We now use these partition functions to construct the characters for the orbifold CFT. 
There are three types of characters to discuss. 
First, 
after gauging $\mathbb{Z}_2$ duality symmetry, 
the $\mathbb{Z}_2$ invariant abelian excitations still remain.
They can fuse with the bosonic $\mathbb{Z}_2$ gauge charge $c$ 
to form $2N$ abelian excitations.
The corresponding characters in the edge CFT are
\begin{align}
\chi_{r+}&=X_{r,r}+\mathcal{Z}_r^{0,\frac{1}{2}},
\nonumber \\
\chi_{r-}&=X_{r,r}-\mathcal{Z}_r^{0,\frac{1}{2}}, 
\end{align}
where $0\leq r<N$. $\psi_{r^-}$ can be understood as $\psi_{r^+}$ attached with a $\mathbb{Z}_2$ charge and satisfies $\psi_{r^-}=\psi_{r^+}\times c$.

Next, non-$\mathbb{Z}_2$-invariant abelian excitations
will combine together to form the $\mathbb{Z}_2$ invariant superselection sector. 
There are $\frac{N(N-1)}{2}$ of them and the corresponding characters are
\begin{align}
\chi_{a,b}=X_{a,b}+X_{b,a}, 
\end{align}
where $0\leq a<b<N$ and the conformal weight for $\chi_{a,b}$ is $h=ab/N$.

The third type of the characters are associated with
the twist fields. 
Although the $\mathbb{Z}_2$ gauge boson is abelian, the $\mathbb{Z}_2$ flux $\sigma$ is non-abelian excitation. 
The bare $\mathbb{Z}_2$ flux $\sigma_0$ can be attached to an abelian excitation, i.e., $\sigma_0\times \psi_{l}=\sigma_{l}$. 
We call the subscript $l$ are the species label of $\sigma$. $\sigma_l$ can fuse with $c$ to form the flux-charge composite particle $\tau_l=\sigma_l\times c$.  The unpaired $\sigma_l$ or $\tau_l$ in the bulk topological phase will leave the $\mathbb{Z}_2$ twist on the boundary and the characters for the $2N$ twist fields are
\begin{align}
\chi_l^{\sigma}&=
\frac{1}{\sqrt{2}}\left[\mathcal{Z}_l^{\frac{1}{2},0}+\mathcal{Z}_l^{\frac{1}{2},\frac{1}{2}}\right],
\nonumber \\
\chi_l^{\tau}&=\frac{1}{\sqrt{2}}\left[\mathcal{Z}_l^{\frac{1}{2},0}-\mathcal{Z}_l^{\frac{1}{2},\frac{1}{2}}\right], 
\end{align}
where $0\leq l<N$.

These characters with their properties are listed in Table \ref{toric_even}. 
\begin{table}[t]
\centering
\begin{ruledtabular}
\begin{tabular}{lccc}
$\chi$ & $d_\chi$ & $h_{\chi}$ & $\mathcal{N}$\\\hline
$\chi_r^+$ & $1$ & $\frac{r^2}{N}$ & $N$\\
$\chi_r^-$ & $1$ & $\frac{r^2}{N}$ & $N$\\
$\chi_{a,b}$ & $2$ & $\frac{ab}{N}$ & $\frac{N(N-1)}{2}$\\
$\chi_l^{\sigma}$ & $\sqrt{N}$ & $-\frac{1}{16}+\frac{l^2}{4N}$ & $N$\\
$\chi_l^{\tau}$ & $\sqrt{N}$ & $-\frac{9}{16}+\frac{l^2}{4N}$ & $N$\\
\end{tabular}
\end{ruledtabular}
\caption{The quantum dimensions $d_\chi$, 
conformal dimensions $h_{\chi}$ (mod $\mathbb{Z}$),
and the number of deconfined fluxes, 
charges and super-sectors
from orbifolding the $\mathbb{Z}_2$ symmetry of 
the $D(\mathbb{Z}_N)$ quantum double model when $N$ is even. 
}\label{toric_even}
\end{table}
\begin{widetext}
The $\mathcal{S}$ for matrix for the characters shown in Table \ref{toric_even} takes this form:
\begin{equation}
\mathcal{S}=\frac{1}{2N}
\begin{pmatrix}
\omega^{2rr^{\prime}} & \omega^{2rr^{\prime}} & 2\omega^{ra^{\prime}+rb^{\prime}} & \sqrt{N}\omega^{rl^{\prime}} & \sqrt{N}\omega^{rl^{\prime}}\\
\omega^{2rr^{\prime}} & \omega^{2rr^{\prime}} & 2\omega^{ra^{\prime}+rb^{\prime}} & -\sqrt{N}\omega^{rl^{\prime}} & -\sqrt{N}\omega^{rl^{\prime}}\\ 
2\omega^{r^{\prime}a+r^{\prime}b} & 2\omega^{r^{\prime}a+r^{\prime}b} & 2[\omega^{ab^{\prime}+ba^{\prime}}+\omega^{aa^{\prime}+bb^{\prime}}] & 0 & 0\\ 
\sqrt{N}\omega^{lr^{\prime}} & -\sqrt{N}\omega^{lr^{\prime}} & 0 & \sqrt{\frac{N}{2}}\omega^{\frac{ll^{\prime}}{2}}[1+(-1)^{\frac{N}{2}+l+l^{\prime}}] & -\sqrt{\frac{N}{2}}\omega^{\frac{ll^{\prime}}{2}}[1+(-1)^{\frac{N}{2}+l+l^{\prime}}]\\
\sqrt{N}\omega^{lr^{\prime}} & -\sqrt{N}\omega^{lr^{\prime}}& 0 & -\sqrt{\frac{N}{2}}\omega^{\frac{ll^{\prime}}{2}}[1+(-1)^{\frac{N}{2}+l+l^{\prime}}] & \sqrt{\frac{N}{2}}\omega^{\frac{ll^{\prime}}{2}}[1+(-1)^{\frac{N}{2}+l+l^{\prime}}]
\end{pmatrix}
\label{Z2even}
\end{equation}
where $\omega=e^{-\frac{2\pi i}{N}}$.
\end{widetext}

Let us now have a close look at 
the simplest case with $N=2$. 
When $N=2$, the $D(\mathbb{Z}_2)$ quantum double model is the toric code model.
After orbifolding the $\mathbb{Z}_2$ symmetry on the edge CFT, it is straightforward to check that the $\mathcal{T}$ and $\mathcal{S}$ matrices are the same as that for non-chiral Ising $\times\ \overline{\mbox{Ising}}$ CFT. Thus the topological phase after gauging $\mathbb{Z}_2$ symmetry has quasiparticle excitations $\{1,\psi,\sigma\}\times \{\overline{1},\overline{\psi},\overline{\sigma}\}$, where $\psi$ is the fermion and $\sigma$ is the non-abelian Ising anyon. They satisfy the following fusion algebra
\begin{align}
\psi\times\psi=1,\quad \psi\times\sigma=\sigma,\quad \sigma\times\sigma=1+\psi. 
\end{align}
The correspondence between $D(\mathbb{Z}_2)/\mathbb{Z}_2$ orbifold CFT and  Ising $\times\ \overline{\mbox{Ising}}$ CFT is shown in Table \ref{quant_double_z2}. $\psi\overline{\psi}$ plays the role of the $\mathbb{Z}_2$ charge and $\overline{\sigma}$ is the $\mathbb{Z}_2$ flux.

\begin{table}[t]
\begin{ruledtabular}
    \begin{tabular}{ccccccccc}
      $\chi_0^+$ & $\chi_1^+$ & $\chi_0^-$ & $\chi_1^-$ & $\chi_{0,1}$ & $\chi_0^{\sigma}$ & $\chi_1^{\sigma}$ & $\chi_0^{\tau}$ & $\chi_1^{\tau}$\\
      \hline
	$1$ & $\psi$ & $\psi\overline{\psi}$ & $\overline{\psi}$ & $\sigma\overline{\sigma}$ & $\overline{\sigma}$ & $\sigma$ & $\overline{\sigma}\psi$ & $\overline{\psi}\sigma$\\
    \end{tabular}
    \end{ruledtabular}
 \caption{
The first row lists the characters for the $D(\mathbb{Z}_2)/\mathbb{Z}_2$ orbifold CFT,
while the second row lists the corresponding primary fields in 
the Ising $\times\ \overline{\mbox{Ising}}$ CFT. For Ising CFT, the three characters are: $\chi_1=(\sqrt{\theta_3/\eta}+\sqrt{\theta_4/\eta})/2$, $\chi_\psi=(\sqrt{\theta_3/\eta}-\sqrt{\theta_4/\eta})/2$ and $\chi_\sigma=\sqrt{\theta_2/\eta}/\sqrt{2}$.
    \label{quant_double_z2}
}
\end{table}

It is also possible to consider 
the reverse process of gauging -- (anyon) condensation. 
In the Ising $\times\ \overline{\mbox{Ising}}$ topological phase, 
after condensing $\mathbb{Z}_2$ boson $\psi\overline{\psi}$, 
$\psi=\overline{\psi}\times \psi \overline{\psi}$ and $\overline{\psi}$ are identified. 
Since the $\mathbb{Z}_2$ fluxes have $-1$ braiding phase with $\psi\overline{\psi}$ 
(This can be seen from the $\mathcal{S}$ matrix), 
they are confined. 
The superselection sector will split into two sectors, which are $e$- and $m$-particles 
in the toric code model.

When $N\geq 4$, 
we can obtain more general $\mathbb{Z}_N$ parafermionic twist liquid described by the $\mathcal{S}$ and $\mathcal{T}$ matrices shown previously.
The fusion algebra can be calculated by using the Verlinde formula.\cite{BarkeshliBondersonChengWang14}  
The details are not shown here. 

Before we move on, we would like to summarize the method we used so far to construct $\mathbb{Z}_2$ orbifold CFT: (1) we first compute the partition function $\mathcal{Z}^{\frac{1}{2},0}$ with twist in the time direction with the zero mode part is  restricted to the $\mathbb{Z}_2$ symmetric mode.  We use these partition functions and the $\mathbb{Z}_2$ symmetric characters in the parent abelian topological phase to construct the characters for abelian Type (1) excitation in the bulk discussed in Sec.~\ref{gauge_anyonic}. (2) we group $\mathbb{Z}_2$ non-symmetric characters to superselection sectors which correspond to non-abelian Type (2) excitations in the twist liquids. (3) By performing modular $S$ transformation for $\mathcal{Z}^{\frac{1}{2},0}$, we can get $\mathcal{Z}^{0,\frac{1}{2}}$ with twist in the spatial direction. Further applying $\mathcal{T}$ transformation upon it will lead to $\mathcal{Z}^{\frac{1}{2},\frac{1}{2}}$. Combining $\mathcal{Z}^{0,\frac{1}{2}}$  and $\mathcal{Z}^{\frac{1}{2},\frac{1}{2}}$ properly, we will obtain the characters for the twist fields, which correspond to type (3) excitations.

\subsubsection{Modify the orbifold CFT by a SPT}
\label{SPT_SET}
If we stack a $\mathbb{Z}_2$ SPT phase discussed in Sec.~\ref{orb_z2_spt} 
on top of the $D(\mathbb{Z}_N)$ quantum double model to form a composite system, 
this SPT phase will not change the quasiparticle properties of $D(\mathbb{Z}_N)$ model. 
However, as shown in the Ref.\ \onlinecite{TeoHughesFradkin15,BarkeshliBondersonChengWang14,EtingofNikshychOstrik10}, 
this SPT phase will change some twist defect $F$-symbols for the composite system.
After gauging the $\mathbb{Z}_2$ symmetry, 
the twist liquid can therefore have different $\mathcal{S}$, 
$\mathcal{T}$ matrices and different fusion rules. The number of these topological phases is characterized by cohomology classes $H^3(Z_2,U(1))=\mathbb{Z}_2$, which also classifies $\mathbb{Z}_2$ SPT phases.
In this section, we will try to understand this phenomenon from 
the point of view of the edge CFT.

For the $D(\mathbb{Z}_N)$ quantum double model with a $\mathbb{Z}_2$ SPT phase stacked on top of it, the total $\mathrm{K}$-matrix is
\begin{align}
{\bf K}=\begin{pmatrix}0 & 1& 0 & 0 \\1& 0& 0 & 0\\0 & 0 & 0& N\\ 0 & 0 & N& 0\end{pmatrix}. 
\end{align}
The edge CFT is described by a four-component boson field with the above $\mathrm{K}$-matrix. 
There are four characters for the edge CFT and the $\mathcal{S}$ and $\mathcal{T}$ matrices are the same as that for 
the $D(\mathbb{Z}_N)$ 
edge CFT. 

The $\mathbb{Z}_2$ symmetry acts on the first two bosonic fields 
as a $\pi$ shift, sending $\phi_1\to\phi_1+\pi$ and $\phi_2\to\phi_2+\pi$. 
The $\mathbb{Z}_2$ orbifold CFT for these two bosonic fields has been discussed in Sec.~\ref{orb_z2_spt}. For $\phi_3$ and $\phi_4$, $\mathbb{Z}_2$ symmetry exchanges these two bosonic fields and the orbifold CFT is in the previous section. After orbifolding $\mathbb{Z}_2$ symmetry for this composite system, the new CFT has the same number of characters. We directly list the character in Table \ref{toric_even_spt} (when $N$ is even) and will give an explanation later.

\begin{table}[t]
\centering
\begin{ruledtabular}
\begin{tabular}{lccc}
$\chi$ & $d_\chi$ & $h_{\chi}$ & $\mathcal{N}$ \\\hline
$\chi_r^+=Z^{0,0}X_r+Z^{0,1}\mathcal{Z}_r^{0\frac{1}{2}}$ & $1$ & $\frac{r^2}{N}$  & $N$\\
$\chi_r^-=Z^{0,0}X_r-Z^{0,1}\mathcal{Z}_r^{0\frac{1}{2}}$ & $1$ & $\frac{r^2}{N}$ & $N$ \\
$\chi_{a,b}=Z^{0,0}(X_{a,b}+X_{b,a})$ & $2$ & $\frac{ab}{N}$ & $\frac{N(N-1)}{2}$ \\
$\chi_l^{\sigma}=\frac{1}{\sqrt{2}}(Z^{1,0}\mathcal{Z}_l^{\frac{1}{2},0}+Z^{1,1}\mathcal{Z}_l^{\frac{1}{2},\frac{1}{2}})$ & $\sqrt{N}$ & $-\frac{5}{16}+\frac{l^2}{4N}$ & $N$\\
$\chi_l^{\tau}=\frac{1}{\sqrt{2}}(Z^{1,0}\mathcal{Z}_l^{\frac{1}{2},0}-Z^{1,1}\mathcal{Z}_l^{\frac{1}{2},\frac{1}{2}})$ & $\sqrt{N}$ & $-\frac{13}{16}+\frac{l^2}{4N}$ & $N$\\
\end{tabular}
\end{ruledtabular}
\caption{The quantum dimensions $d_\chi$, 
conformal dimensions $h_{\chi}$ (mod $\mathbb{Z}$) 
and the number of deconfined fluxes, charges and super-sectors 
from orbifolding the $\mathbb{Z}_2$ symmetry of edge CFT of
the $(\mbox{SPT}+D(\mathbb{Z}_N))/\mathbb{Z}_2$ quantum double model with even $N$. 
}\label{toric_even_spt}
\end{table}

In Sec.\ \ref{orb_z2_spt}, 
we showed that for the $\mathbb{Z}_2$ SPT phase, 
after gauging $\mathbb{Z}_2$ symmetry, 
the new topological phase has four quasiparticles $1, e, m, em$, 
where $e$ is the bosonic $\mathbb{Z}_2$ charge and $m$ is the semionic $\mathbb{Z}_2$ flux. 
For the $D(\mathbb{Z}_N)/\mathbb{Z}_2$ topological phase, 
there is also a bosonic $\mathbb{Z}_2$ charge $c$ and $\mathbb{Z}_2$ flux $\sigma$. 
If we put the SPT phase on top of the $D(\mathbb{Z}_N)$ quantum double model, after gauging the $\mathbb{Z}_2$ symmetry, the new $\mathbb{Z}_2$ charge $c^{\prime}=e\times c$ and flux $\sigma^{\prime}=\sigma\times m$ are the bound states of the original $\mathbb{Z}_2$ charges and fluxes. We show the complete table of characters in  Table \ref{toric_even_spt}.  The conformal dimensions of the twist fields are shifted by $-1/4$ which actually is the conformal dimension for the $\mathbb{Z}_2$ flux $m$.

To obtain the  $\mathcal{S}$ matrix for the $(\mbox{SPT}+D(\mathbb{Z}_N))/\mathbb{Z}_2$ CFT, we only need to replace all the $[1+(-1)^{\frac{N}{2}+l+l^{\prime}}]$ in Table \ref{Z2even} to $[1-(-1)^{\frac{N}{2}+l+l^{\prime}}]$, which is also the $\pi$ monodromy phase between a pair of semions.  
This sign difference comes from the phase factor
acquired by $Z^{k,l}$ in the $S$ transformation
\begin{align}
Z^{k,l}(-1/\tau)=(-1)^{kl}Z^{l,k}(\tau), 
\end{align}
where $l,k=0,1$. 
The fusion rule for this $\mathcal{S}$ matrix is the same that for Table \ref{Z2even}.

\subsubsection{Orbifolding $\mathbb{Z}_2$ symmetry when $N$ is odd}
When $N$ is odd, the calculation is similar to the even $N$ case. However, in this case, the bare twist field already carries $1/2$ species label. We directly write down the characters without too much explanation.

The characters for twist fields are given by 
\begin{widetext}
\begin{align}
\chi_{l}^{\sigma}&=
\frac{1}{\sqrt{2}\eta(\tau)}\left(\Theta_{n,8N}+\Theta_{n+2N, 8N}\right.
+\left.\Theta_{n+4N,8N}+\Theta_{n+6N,8N}\right)\sqrt{\frac{\overline{\eta}}{\overline{\theta}_4}}
\nonumber \\
&\quad 
+\frac{1}{\sqrt{2}\eta(\tau)}\left(\Theta_{n,8N}+e^{-2\pi i\frac{N+n}{4}}\Theta_{n+2N, 8N}\right.
-\left.\Theta_{n+4N,8N}-e^{2\pi i\frac{N+n}{4}}\Theta_{n+6N,8N}\right)\sqrt{\frac{\overline{\eta}}{\overline{\theta}_3}},
\nonumber \\
\chi_{l}^{\tau}&=
\frac{1}{\sqrt{2}\eta(\tau)}\left(\Theta_{n,8N}+\Theta_{n+2N, 8N}\right.
+\left.\Theta_{n+4N,8N}+\Theta_{n+6N,8N}\right)\sqrt{\frac{\overline{\eta}}{\overline{\theta}_4}}
\nonumber \\
&\quad -\frac{1}{\sqrt{2}\eta(\tau)}\left(\Theta_{n,8N}+e^{-2\pi i\frac{N+n}{4}}\Theta_{n+2N, 8N}\right.
-\left.\Theta_{n+4N,8N}-e^{2\pi i\frac{N+n}{4}}\Theta_{n+6N,8N}\right)\sqrt{\frac{\overline{\eta}}{\overline{\theta}_3}},
\end{align}
\end{widetext}
where $0\leq l<N$ and $n=2l+1$.

The original $\mathbb{Z}_2$ invariant characters before gauging will split into two sectors, which differ by the $\mathbb{Z}_2$ charge,
as 
\begin{align}
\chi_r^{\pm}&=\mathcal{Z}_{r,r}^{0,0}\pm \frac{1}{\eta(\tau)}
\left[\Theta_{4r,8N}(\tau)-\Theta_{4r+4N,8N}
\right] 
\sqrt{\frac{\eta(\bar{\tau})}{\theta_2(\bar{\tau})}},
%
\end{align}
where $0\leq r<N$.

The original characters which are not invariant under $\mathbb{Z}_2$ symmetry will form the superselection sectors:
\begin{equation}
\chi_{a,b}=\mathcal{Z}_{a,b}+\mathcal{Z}_{b,a}
\end{equation}
where $0\leq a<b<N$.

The characters with their properties are listed in 
Table \ref{toric_odd}.
\begin{table}[t]
\centering
\begin{ruledtabular}
\begin{tabular}{lccc}
$\chi$ & $d_\chi$ & $h_{\chi}$ & $\mathcal{N}$\\\hline
$\chi_r^+$ & $1$ & $\frac{r^2}{N}$ & $N$\\
$\chi_r^-$ & $1$ & $\frac{r^2}{N}$ & $N$\\
$\chi_{a,b}$ & $2$ & $\frac{ab}{N}$ & $\frac{N(N-1)}{2}$\\
$\chi_l^{\sigma}$ & $\sqrt{N}$ & $-\frac{1}{16}+\frac{(l+\frac{1}{2})^2}{4N}$ & $N$\\
$\chi_l^{\tau}$ & $\sqrt{N}$ & $-\frac{9}{16}+\frac{(l+\frac{1}{2})^2}{4N}$ & $N$\\
\end{tabular}
\end{ruledtabular}
\caption{The quantum dimensions $d_\chi$, conformal dimension $h_{\chi}$ and the number of deconfined fluxes, charges and super-sectors from orbifolding the $\mathbb{Z}_2$ symmetry of the edge of $D(\mathbb{Z}_N)$ quantum double model with odd $N$. 
}\label{toric_odd}
\end{table}

\begin{table}[t]
  \begin{center}
  \begin{ruledtabular}
    \begin{tabular}{cccccccccc}
      ab & 00 & 01 & 02 & 10 & 11 & 12 & 20 & 21 & 22\\
      \hline
	$\lambda\bar{\lambda}$ & $0\bar{0}$ & $1\bar{1}$ & $2\bar{2}$ & $1\bar{2}$ & $2\bar{0}$ & $0\bar{1}$ & $2\bar{1}$ & $0\bar{2}$ & $1\bar{0}$\\
    \end{tabular}
 \caption{
The first row lists the quasiparticle excitations 
for the $D(\mathbb{Z}_3)$ quantum double model,
which are labeled by the two-component vector $(a,b)$ with $0\leq a,b<N$.
The second row lists the corresponding quasiparticle excitations for the $SU(3)_1\times\overline{SU(3)_1}$ topological phase. 
The chiral $SU(3)_1$ topological phase has three excitations 
(defined in the next section) and are labeled by $\lambda=0,1,2$. 
The excitations in the anti-chiral $\overline{SU(3)_1}$ phase 
are labeled by $\bar{\lambda}={\bar{0},\bar{1},\bar{2}}$.
    \label{quant_double_z3}
}
\end{ruledtabular}
  \end{center}
\end{table}

\begin{widetext}
The $\mathcal{S}$ matrix for the corresponding characters listed in Table \ref{toric_odd} takes this form:
\begin{equation}
\mathcal{S}=\frac{1}{2N}
\begin{pmatrix}
\omega^{2rr^{\prime}} & \omega^{2rr^{\prime}} & 2\omega^{ra^{\prime}+rb^{\prime}} & \sqrt{N}\omega^{\frac{rn^{\prime}}{2}} & \sqrt{N}\omega^{\frac{rn^{\prime}}{2}}\\
\omega^{2rr^{\prime}} & \omega^{2rr^{\prime}} & 2\omega^{ra^{\prime}+rb^{\prime}} & -\sqrt{N}\omega^{\frac{rn^{\prime}}{2}} & -\sqrt{N}\omega^{\frac{rn^{\prime}}{2}}\\ 
2\omega^{r^{\prime}a+r^{\prime}b} & 2\omega^{r^{\prime}a+r^{\prime}b} & 2[\omega^{ab^{\prime}+ba^{\prime}}+\omega^{aa^{\prime}+bb^{\prime}}] & 0 & 0\\ 
\sqrt{N}\omega^{\frac{nr^{\prime}}{2}} & -\sqrt{N}\omega^{\frac{ar^{\prime}}{2}} & 0 & \sqrt{\frac{N}{2}}\omega^{\frac{nn^{\prime}}{8}}[1+(i)^{-(N+n+n^{\prime})}] & -\sqrt{\frac{N}{2}}\omega^{\frac{nn^{\prime}}{8}}[1+(i)^{-(N+n+n^{\prime})}]\\
\sqrt{N}\omega^{\frac{nr^{\prime}}{2}} & -\sqrt{N}\omega^{\frac{nr^{\prime}}{2}}& 0 & -\sqrt{\frac{N}{2}}\omega^{\frac{nn^{\prime}}{8}}[1+(i)^{-(N+n+n^{\prime})}] & \sqrt{\frac{N}{2}}\omega^{\frac{nn^{\prime}}{8}}[1+(i)^{-(N+n+n^{\prime})}]
\end{pmatrix}
\label{Z2odd}
\end{equation}
\end{widetext}
where $\omega=e^{-\frac{2\pi i}{N}}$.

Let us now discuss the cases of $N=1$ and $N=3$ in detail. 
When $N=1$, before orbifolding, there is no topological order in the $2+1$-dimensional bulk phase. 
The orbifold edge CFT has four primary fields and the $\mathcal{S}$ and $\mathcal{T}$ matrices 
are the same as that for the toric code model. Notice that the $\mathbb{Z}_2$ symmetry we are considering here is different from that discussed in Eq.\eqref{z2_symm}.
According to our discussion in Sec.~\ref{orb_z2_spt}, 
the bulk phase before gauging is a trivial SPT phase.

When $N=3$, the $D(\mathbb{Z}_3)/\mathbb{Z}_2$ orbifold CFT has 15 primary fields. 
Before gauging, the parent $D(\mathbb{Z}_3)$ quantum double model 
is equivalent to non-chiral $SU(3)_1\times\overline{SU(3)_1}$ topological phase through a $SL(2,\mathbb{Z})$ similarity transformation. 
Their correspondence is shown in Table \ref{quant_double_z3}. 
The $SU(3)_1$ topological phase is described by the Cartan $\mathrm{K}$-matrix in Eq.\ \eqref{K_su3} 
and will be explained in the next section. 
The $\mathbb{Z}_2$ duality symmetry in $D(\mathbb{Z}_3)$ topological phase is equivalent to $\mathbb{Z}_2$ bilayer/charge conjugation symmetry defined in the chiral sector of $SU(3)_1\times\overline{SU(3)_1}$ topological phase. 
Since $SU(3)_1/\mathbb{Z}_2=SU(2)_4$ CFT (discussed in the next section), 
the edge CFT of $D(\mathbb{Z}_3)$ quantum double model after orbifolding $\mathbb{Z}_2$ symmetry is the same as $SU(2)_4\times \overline{SU(3)_1}$ CFT.

%
%
%

\subsection{Orbifolding $\mathbb{Z}_2$ symmetry in $SU(3)_1$ CFT}

The $SU(3)_1$ topological phase is a bosonic bilayer quantum Hall state 
with the {K}-matrix 
\begin{equation}
\textbf{K}=\begin{pmatrix}2&-1\\-1&2 \end{pmatrix}. 
\label{K_su3}
\end{equation}
This is an abelian topological phase with three quasi-particle excitation $\psi_0=(0,0), \psi_1=(0,1)$ and $\psi_2=(1,0)$. The fusion rules for these three abelian anyons are $\psi_1\times \psi_1=\psi_2$, $\psi_1\times \psi_2=\psi_0$. The edge theory for this topological phase is described the chiral $SU(3)_1$ CFT. This CFT is equivalent to the chiral Luttinger liquid 
with the $\mathrm{K}$-matrix defined in Eq.\ \eqref{K_su3} and the corresponding three characters are $Z_{0,0}$, $Z_{0,1}$ and $Z_{1,0}$,
\begin{equation}
Z_{\psi_i}=\left(\frac{1}{\eta}\right)^2 \sum_{\vec{\Lambda}}q^{\frac{1}{2}(\textbf{K}\vec{\Lambda}+\psi_i)^T\textbf{K}^{-1}(\textbf{K}\vec{\Lambda}+\psi_i)}
\end{equation}
where $\vec{\Lambda}$ are three two-component integer-valued vectors and $i=0,1,2$.

This model has an effective $\mathbb{Z}_2$ bilayer symmetry which is represented by the matrix $M_1=\sigma_x$ and exchanges $\psi_1$ and $\psi_2$. This model also has charge conjugation symmetry and is described by $M_2=-I$, which exchanges the anyon class $\psi_1$ and $\psi_2$ and thus is equivalent to the $\mathbb{Z}_2$ bilayer symmetry. After gauging $M_1$ or $M_2$ symmetry, the twist liquid will have five different anyon excitations. The original $SU(3)_1$ vacuum sector splits into two sectors differed by a $\mathbb{Z}_2$ charge $c$. The $\mathbb{Z}_2$ non-invariant sectors $\psi_1$ and $\psi_2$ will combine together to form the superselection sector $\psi\equiv[\psi_1+\psi_2]$. The $\mathbb{Z}_2$ flux $\sigma$ does not have species labels and attaching a $\mathbb{Z}_2$ charge to it will form a composite flux-charge particle $\tau$. 

Gauging $M_1$ and $M_2$ symmetries will lead to the same twist liquid. However, on the edge CFT, this corresponds to two different orbifolding procedures. We will show later that after gauging, they have the same $\mathcal{S}$ and $\mathcal{T}$ matrices. Here we will orbifold $\mathbb{Z}_2$ bilayer symmetry for $SU(3)_1$ CFT first.

\subsubsection{Orbifolding $\mathbb{Z}_2$ bilayer symmetry}
Under the $\mathbb{Z}_2$ bilayer symmetry operator $M_1$, $\phi_1\leftrightarrow \phi_2$. In the $\vec{\varphi}$ basis, $\varphi^1\to\varphi^1$ and $\varphi^2\to-\varphi^2$. Thus we need to consider the twist boundary condition for $\varphi_2$ field in both time and spatial directions. The calculation is similar to that for $D(\mathbb{Z}_N)/\mathbb{Z}_2$ CFT in the previous section. We directly list the characters of $SU(3)_1/M_1$ in the Table \ref{su3_1}.

\begin{table}[t]
\centering
\begin{ruledtabular}
\begin{tabular}{lll}
character $\chi$ & $d_\chi$ & $h_\chi$ \\\hline
$\chi_{\mathbb{I}}=Z_{0,0}+\frac{1}{\eta}(\Theta_{0,8}-\Theta_{4,8})\sqrt{\frac{\eta}{\theta_2}}$ & $1$ & $0$\\
$\chi_{c}=Z_{0,0}-\frac{1}{\eta}(\Theta_{0,8}-\Theta_{4,8})\sqrt{\frac{\eta}{\theta_2}}$ & $1$ & $0$\\
$\chi_{\psi}=Z_{1,0}+Z_{0,1}$ & $2$ & $1/3$\\
$\begin{array}{cc}\chi_{\sigma}&=\frac{1}{\sqrt{2}}\frac{1}{\eta}(\Theta_{1,8}+\Theta_{3,8}+\Theta_{5,8}+\Theta_{7,8})\sqrt{\frac{\eta}{\theta_4}}\\ &+\frac{1}{\sqrt{2}}\frac{1}{\eta}(\Theta_{1,8}-\Theta_{3,8}-\Theta_{5,8}+\Theta_{7,8})\sqrt{\frac{\eta}{\theta_3}}\end{array}$ & $\sqrt{3}$ & $1/8$\\
$\begin{array}{cc}\chi_{\tau}&=\frac{1}{\sqrt{2}}\frac{1}{\eta}(\Theta_{1,8}+\Theta_{3,8}+\Theta_{5,8}+\Theta_{7,8})\sqrt{\frac{\eta}{\theta_4}}\\ &-\frac{1}{\sqrt{2}}\frac{1}{\eta}(\Theta_{1,8}-\Theta_{3,8}-\Theta_{5,8}+\Theta_{7,8})\sqrt{\frac{\eta}{\theta_3}}\end{array}$ & $\sqrt{3}$ & $5/8$\\
\end{tabular}
\end{ruledtabular}
\caption{The quantum dimensions $d_\chi$ and spin statistics $\theta_\chi=e^{2\pi ih_\chi}$ of deconfined fluxes, charges and super-sectors from orbifolding the $\mathbb{Z}_2$ bilayer symmetry of $SU(3)_1$. 
}\label{su3_1}
\end{table}

The $\mathcal{S}$ matrix for the characters is
\begin{equation}
\mathcal{S}=\frac{1}{2\sqrt{3}}\begin{pmatrix}
1&1&2&\sqrt{3}&\sqrt{3}\\
1&1&2&-\sqrt{3}&-\sqrt{3}\\
2&2&-2&0&0\\
\sqrt{3}&-\sqrt{3}&0&\sqrt{3}&-\sqrt{3}\\
\sqrt{3}&-\sqrt{3}&0&-\sqrt{3}&\sqrt{3}
\end{pmatrix}.
\end{equation}

\subsubsection{Orbifolding $\mathbb{Z}_2$ charge conjugation symmetry}
Under the charge conjugation symmetry operator $M_2$, $\phi_1\to -\phi_1$ and $\phi_2\to -\phi_2$. This means that in the $\vec{\varphi}$ basis, $\varphi^{i}\to -\varphi^i$ ($i=1,2$). It is straightforward to calculate the partition function with this twist boundary condition and we list all the characters in Table \ref{su3_2}.

\begin{table}[t]
\centering
\begin{ruledtabular}
\begin{tabular}{lll}
character $\chi$ & $d_\chi$ & $h_\chi$
\\\hline
$\chi_{\mathbb{I}}=Z_{0,0}+\frac{\eta}{\theta_2}$ & $1$ & $0$\\
$\chi_{c}=Z_{0,0}-\frac{\eta}{\theta_2}$ & $1$ & $0$\\
$\chi_{\psi}=Z_{1,0}+Z_{0,1}$ & $2$ & $1/3$\\
$\chi_{\sigma}=\frac{\eta}{\theta_4}+\frac{\eta}{\theta_3}$ & $\sqrt{3}$ & $1/8$\\
$\chi_{\tau}=\frac{\eta}{\theta_4}-\frac{\eta}{\theta_3}$ & $\sqrt{3}$ & $5/8$\\
\end{tabular}
\end{ruledtabular}
\caption{The quantum dimensions $d_\chi$ and spin statistics $\theta_\chi=e^{2\pi ih_\chi}$ of deconfined fluxes, charges and super-sectors from orbifolding the $\mathbb{Z}_2$ charge conjugation symmetry of $SU(3)_1$. 
}\label{su3_2}
\end{table}

It is interesting to notice that $SU(3)_1/\mathbb{Z}_2$ CFT has the same $\mathcal{S}$ and $\mathcal{T}$ matrices as the $SU(2)_4$ CFT. $SU(2)_4$ chiral CFT has five primary fields labeled by $j={\bf 0, 1/2, 1, 3/2, 2}$ with conformal dimensions
\begin{align}
h_j=\frac{j(j+1)}{6}=0, \frac{1}{8},\frac{1}{3},\frac{5}{8},1
\end{align}
and the $\mathcal{S}$-matrix 
\begin{align}
\mathcal{S}_{j_1,j_2}=\frac{1}{\sqrt{3}}\sin\left[\frac{\pi (2j_1+1)(2j_2+1)}{6}\right]
\end{align}

The primary field ${\bf 0}$ serves as the vacuum, ${\bf 2}$ is the boson $\mathbb{Z}_2$ charge $c$ in $SU(3)_1/\mathbb{Z}_2$ CFT, ${\bf 1}$ corresponds to the superselection sector $\psi$, and ${\bf 1/2}$, ${\bf 3/2}$ are identified as the twist fields $\sigma$ and $\tau$. The primary fields satisfy the fusion algebra
\begin{align}
&
{\bf 2\times 2=0},\quad {\bf 2\times 1=1}, \quad{\bf 2\times \frac{1}{2}=\frac{3}{2}}\nonumber\\
&
{\bf \frac{1}{2}\times \frac{1}{2}=0+1},\quad {\bf \frac{1}{2}\times 1=\frac{1}{2}+\frac{3}{2}}\nonumber\\
&
{\bf 1\times 1=0+1+2}
\end{align}

\subsection{Edge CFT of $SO(8)_1$ state}
The $SO(8)_1$ bosonic Abelian topological phase is described by a Chern-Simons theory with
\begin{equation}
\textbf{K}=\begin{pmatrix} 2 & -1&-1&-1 \\ -1&2&0&0\\-1 &0&2&0\\-1&0&0&2 \end{pmatrix}.
\label{K_so8}
\end{equation}
The ${K}$-matrix is identical to the Cartan matrix of the Lie algebra $so(8)$ and as a result, the edge CFT carries a chiral $SO(8)$ Kac-Moody structure at level 1. The bulk topological theory has four quasi-particle excitation $\psi_{\lambda_0}$, $\psi_{\lambda_1}$, $\psi_{\lambda_2}$ and $\psi_{\lambda_3}$ where $\vec{\lambda}_0^T=(0,0,0,0)$, $\vec{\lambda}_1^T=(0,1,0,0)$, $\vec{\lambda}_2^T=(0,0,1,0)$ and $\vec{\lambda}_3^T=(0,0,0,1)$. $\psi_{\lambda_0}$ is vacuum and $\psi_{\lambda_i}$ ($i=1,2,3$) are all fermions with mutual semionic statistics ($\mathcal{D}\mathcal{S}_{ij}=-1$ ($i,j=1,2,3$)). The fermions obey
\begin{align}
\psi_i^2=1,\quad\psi_1\times\psi_2\times\psi_3=1
\end{align}
This model has a triality anyonic symmetry $S_3$,
which is generated by a threefold rotation $\rho$ and a twofold reflection $\sigma_1$.\cite{khan2014, BarkeshliBondersonChengWang14, TeoHughesFradkin15} $\rho$ cyclicly rotates $\psi_i\to\psi_{i+1}$ and $\sigma_1$ exchanges $\psi_{2}$ and $\psi_{3}$ while fixes $\psi_{1}$. The other two reflection operators are defined as
\begin{align}
\sigma_2=\sigma_1\rho,\quad \sigma_3=\sigma_1\rho^2
\end{align}
They fix $\psi_2$ and $\psi_3$ respectively while interchange the other two fermions. 
The edge CFT is described by chiral Luttinger liquid with 
the $\mathrm{K}$-matrix defined in Eq.\ \eqref{K_so8}. 
This CFT has central charge $c=4$ and has four characters corresponding to four different quasi-particle sectors in the bulk
\begin{equation}
\chi_{\lambda_i}=\left(\frac{1}{\eta}\right)^4 \sum_{\vec{\Lambda}}q^{\frac{1}{2}(\textbf{K}\vec{\Lambda}+\vec{\lambda_i})^T\textbf{K}^{-1}(\textbf{K}\vec{\Lambda}+\vec{\lambda_i})}
\end{equation}
where $\vec{\Lambda}$ are the four component integer-valued vectors.

\subsubsection{Orbifolding $\mathbb{Z}_3$ symmetry}
After orbifolding the $\mathbb{Z}_3$ symmetry, the $\mathbb{Z}_3$ invariant vacuum state $\psi_{\lambda_0}$ will split into three vacuums $\mathrm{I}, z_3,\overline{z}_3$ and they are differed by the $\mathbb{Z}_3$ gauge charge. $\psi_{\lambda_i}$ with $i=1,2,3$ are not invariant under $\mathbb{Z}_3$ rotation and will form the superselection sector $[\psi_{\lambda_1}+\psi_{\lambda_2}+\psi_{\lambda_3}]$. The  $\mathbb{Z}_3$ gauge flux $\rho_0$ does not have species labels for it (the quotient $\mathcal{A}_{SO(8)_1}/(1-\Lambda_3)\mathcal{A}_{SO(8)_1}$ is trivial). $\rho_0$ can fuse with the $\mathbb{Z}_3$ charge to get $\rho_1=\rho_0\times z_3$ and $\rho_2=\rho_0\times \overline{z}_3$. Similarly, the anti-particles of $\mathbb{Z}_3$ flux are $\overline{\rho}_0$, $\overline{\rho}_1$ and $\overline{\rho}_2$. 

Having a physical picture in mind, we now explicitly calculate the characters for the orbifold CFT.

The threefold rotation symmetry operator $\rho$ is defined as 
\begin{equation}
\rho=\begin{pmatrix} 1 & 0&0&0 \\ 0&0&1&0\\0 &0&0&1\\0&1&0&0 \end{pmatrix}
\label{3fold_so8}
\end{equation}
which satisfies $\rho\textbf{K}\rho^T=\textbf{K}$. For the four-component bosonic field $\vec{\Phi}$ living on the boundary, under $\rho$, $\phi^1\to\phi^1,\phi^2\to\phi^3,\phi^3\to\phi^4,\phi^4\to\phi^2$. We can simultaneously diagonalize ${\bf K}$ and $\rho$ by defining new bosonic field $\vec{\varphi}$, so that under the threefold rotation, $\varphi^2\to\omega^{-1}\varphi^2,\varphi^3\to\omega\varphi^3, \varphi^4\to\varphi^4$, where $\omega=e^{2\pi i/3}$. Thus in the language of partition function, $\rho$ will generate $\mathbb{Z}_3$ twist for $\varphi_2$ and $\varphi_3$ in both the spatial and time directions.

Here we will first consider the partition function with twist boundary condition in the time direction. After orbifolding the $\mathbb{Z}_3$ symmetry, only the state invariant under $\rho$ will survive, i.e., $\rho|\textbf{K}\vec{\Lambda}+\vec{\lambda}\rangle=|\textbf{K}\vec{\Lambda}+\vec{\lambda}\rangle$. This requires that $\vec{\Lambda}^T=(n,m,m,m)$ and $\vec{\lambda}=\vec{\lambda}_0$. The partition function with twist boundary condition in the time direction is
\begin{eqnarray}
\nonumber \mathcal{Z}^{0,\nu} &=&\left(\frac{1}{\eta}\right)^2\sum_{\vec{\Lambda}_{\rho}}q^{\frac{1}{2}\vec{\Lambda}_{\rho}^T\textbf{K}_{\rho}\vec{\Lambda}_{\rho}}\frac{\eta(\tau)}{\theta_{\beta}^{\frac{1}{2}}(\tau)}e^{2\pi i\frac{1}{2}\beta}\\
&=&B_0Z^{0,\nu}
\end{eqnarray}
where $\nu=0,\frac{1}{3},\frac{2}{3}$ and $\theta_{\beta}^{\frac{1}{2}}(\tau)$
is defined in Eq.\, \eqref{theta} with $\beta=\frac{1}{2}+\nu$, $\vec{\Lambda}_{\rho}^T=(m,n)$  and  
\begin{equation}
\textbf{K}_{\rho}=\begin{pmatrix} 2 & -3\\-3&6\end{pmatrix}
\end{equation}
is the $\mathrm{K}$-matrix when projected to the two-dimensional $\mathbb{Z}_3$ symmetric lattice $\Lambda=(n,m,m,m)$.
The three characters for this $\mathrm{K}$-matrix are $B_0, B_1$ and $B_2$.
Under the $T$ transformation,
\begin{align}
B_0(\tau+1)&=B_0(\tau), 
\nonumber \\
B_1(\tau+1)&=\omega B_1(\tau), 
\nonumber \\
B_2(\tau+1)&=\omega B_2(\tau), 
\end{align}
where $\omega=e^{2\pi i/3}$.
Under the $S$ transformation,
\begin{align}
B_0(-{1}/{\tau})&=\frac{1}{\sqrt{3}}(B_0+B_1+B_2),
\nonumber \\
B_1(-{1}/{\tau})&=\frac{1}{\sqrt{3}}(B_0+\omega B_1+\omega^2 B_2),
\nonumber \\
B_2(-{1}/{\tau})&=\frac{1}{\sqrt{3}}(B_0+\omega^2 B_1+\omega B_2).
\end{align}

The partition function with twist boundary condition in the space direction is obtained by $S$ transformation from the $\mathcal{Z}^{0,\nu}$
\begin{eqnarray}
\nonumber \mathcal{Z}^{\mu,0}&=&\frac{1}{\sqrt{3}}(B_0+B_1+B_2)\frac{\eta(\tau)}{\theta_{\frac{1}{2}}^{\alpha}(\tau)}e^{2\pi i\alpha\frac{1}{2}}\\
&=&\frac{1}{\sqrt{3}}(B_0+B_1+B_2)Z^{\mu,0}
\end{eqnarray}
where $\mu=0,\frac{1}{3},\frac{2}{3}$ and $\alpha=\frac{1}{2}-\mu$.

Similarly, 
the partition functions for the other twisted sectors 
$\mathcal{Z}^{\mu,\nu}$ can also be calculated:
\begin{eqnarray}
\nonumber &&\mathcal{Z}^{\frac{2}{3},\frac{1}{3}}=\frac{1}{\sqrt{3}}(B_0+\omega
             B_1+\omega B_2)Z^{\frac{2}{3},\frac{1}{3}},
  \\
\nonumber
          &&\mathcal{Z}^{\frac{2}{3},\frac{2}{3}}=\frac{1}{\sqrt{3}}(B_0+\omega^2 B_1+\omega^2 B_2)Z^{\frac{2}{3},\frac{2}{3}},
  \\
\nonumber
          &&\mathcal{Z}^{\frac{1}{3},\frac{1}{3}}=\frac{1}{\sqrt{3}}(B_0+\omega^2 B_1+\omega^2 B_2)Z^{\frac{1}{3},\frac{1}{3}},
  \\
\nonumber &&\mathcal{Z}^{\frac{1}{3},\frac{2}{3}}=\frac{2}{\sqrt{3}}(B_0+\omega B_1+\omega B_2)Z^{\frac{1}{3},\frac{2}{3}},
\end{eqnarray}
where $Z^{\mu,\nu}=\frac{\eta(\tau)}{\theta^{\alpha}_{\beta}(\tau)}e^{2\pi
  i\alpha\beta}$.  $\theta^{\alpha}_{\beta}(\tau)$ is defined in
Eq.\ \eqref{theta} and their modular transformation properties are listed in Eq.\ \eqref{thetast}, which are quite useful in the calculation of the $\mathcal{S}$ and $\mathcal{T}$ matrices for the characters. 

From the partition function obtained for different sectors, we can construct the characters of $SO(8)_1/\mathbb{Z}_3$ CFT. The results are listed in the Table. \ref{so8z3}.

\begin{table}[t]
\centering
\begin{ruledtabular}
\begin{tabular}{lll}
character $\chi$ & $d_\chi$ & $h_\chi$ \\\hline
$\chi_{\mathbb{I}}=\chi_{\lambda_0}+e^{\frac{2\pi i}{12}}\mathcal{Z}^{0,\frac{1}{3}}+e^{-\frac{2\pi i}{12}}\mathcal{Z}^{0,\frac{2}{3}}$ & $1$ & $0$\\
$\chi_{z_3}=\chi_{\lambda_0}+\omega^{-1} e^{\frac{2\pi i}{12}}\mathcal{Z}^{0,\frac{1}{3}}+\omega e^{-\frac{2\pi i}{12}}\mathcal{Z}^{0,\frac{2}{3}}$ & $1$ & $0$\\
$\chi_{\overline{z}_3}=\chi_{\lambda_0}+\omega e^{\frac{2\pi i}{12}}\mathcal{Z}^{0,\frac{1}{3}}+\omega^{-1} e^{-\frac{2\pi i}{12}}\mathcal{Z}^{0,\frac{2}{3}}$ & $2$ & $0$\\
$\chi_{\psi}=\chi_{\lambda_1}+\chi_{\lambda_2}+\chi_{\lambda_3}$ & $3$ & $1/2$\\
$\chi_{\rho_0}=\mathcal{Z}^{\frac{2}{3},0}+\mathcal{Z}^{\frac{2}{3},\frac{1}{3}}+\mathcal{Z}^{\frac{2}{3},\frac{2}{3}}$ & 2 & $1/9$\\
$\chi_{\rho_1}=\mathcal{Z}^{\frac{2}{3},0}+\omega^{-1}\mathcal{Z}^{\frac{2}{3},\frac{1}{3}}+\omega\mathcal{Z}^{\frac{2}{3},\frac{2}{3}}$ & 2 & $4/9$\\
$\chi_{\rho_2}=\mathcal{Z}^{\frac{2}{3},0}+\omega\mathcal{Z}^{\frac{2}{3},\frac{1}{3}}+\omega^{-1}\mathcal{Z}^{\frac{2}{3},\frac{2}{3}}$ & 2 & $7/9$\\
$\chi_{\overline{\rho}_0}=\mathcal{Z}^{\frac{1}{3},0}+\mathcal{Z}^{\frac{1}{3},\frac{1}{3}}+\mathcal{Z}^{\frac{1}{3},\frac{2}{3}}$ & 2 & $1/9$\\
$\chi_{\overline{\rho}_1}=\mathcal{Z}^{\frac{1}{3},0}+\omega^{-1}\mathcal{Z}^{\frac{1}{3},\frac{1}{3}}+\omega\mathcal{Z}^{\frac{1}{3},\frac{2}{3}}$ & 2 & $4/9$\\
$\chi_{\overline{\rho}_2}=\mathcal{Z}^{\frac{1}{3},0}+\omega\mathcal{Z}^{\frac{1}{3},\frac{1}{3}}+\omega^{-1}\mathcal{Z}^{\frac{1}{3},\frac{2}{3}}$ & 2 & $7/9$\\
\end{tabular}
\end{ruledtabular}
\caption{The quantum dimensions $d_\chi$ and spin statistics $\theta_\chi=e^{2\pi ih_\chi}$ of deconfined fluxes, charges and super-sectors from orbifolding the $\mathbb{Z}_3$ symmetry of $so(8)_1$. 
}\label{so8z3}
\end{table}

\begin{widetext}
The $\mathcal{S}$ matrix for the corresponding characters (in the Table. \ref{so8z3}) in $\mathbb{Z}_3$ orbifold theory is
\begin{align}
\mathcal{S}=\frac{1}{\mathcal{D}}
\begin{pmatrix}
 1 & 1 & 1 & 3 & 2 & 2 & 2 & 2 & 2 & 2 \\
1 & 1 & 1 & 3 & 2\omega^2 & 2\omega^2 & 2\omega^2 & 2\omega & 2\omega & 2\omega \\
1 & 1 & 1 & 3 & 2\omega & 2\omega & 2\omega & 2\omega^2 & 2\omega^2 & 2\omega^2 \\
 3 & 3 & 3 & -3 & 0 & 0 & 0& 0& 0& 0 \\
 2 & 2\omega^2 & 2\omega & 0& 2e^{-\frac{13}{18}2\pi i} & 2\omega^2 e^{-\frac{13}{18}2\pi i} & 2\omega e^{-\frac{13}{18}2\pi i} & 2e^{\frac{13}{18}2\pi i} & 2\omega e^{\frac{13}{18}2\pi i} &2\omega^2 e^{\frac{13}{18}2\pi i}\\
2 & 2\omega^2 & 2\omega & 0& 2\omega^2 e^{-\frac{13}{18}2\pi i} & 2\omega e^{-\frac{13}{18}2\pi i} & 2 e^{-\frac{13}{18}2\pi i} & 2\omega e^{\frac{13}{18}2\pi i} & 2\omega^2 e^{\frac{13}{18}2\pi i} &2 e^{\frac{13}{18}2\pi i}\\
2 & 2\omega^2 & 2\omega & 0& 2\omega e^{-\frac{13}{18}2\pi i} & 2 e^{-\frac{13}{18}2\pi i} & 2\omega^2 e^{-\frac{13}{18}2\pi i} & 2\omega^2 e^{\frac{13}{18}2\pi i} & 2 e^{\frac{13}{18}2\pi i} &2\omega e^{\frac{13}{18}2\pi i}\\
2 & 2\omega & 2\omega^2 & 0& 2 e^{\frac{13}{18}2\pi i} & 2\omega e^{\frac{13}{18}2\pi i} & 2\omega^2 e^{\frac{13}{18}2\pi i} & 2 e^{-\frac{13}{18}2\pi i} & 2\omega^2 e^{-\frac{13}{18}2\pi i} &2\omega e^{-\frac{13}{18}2\pi i}\\
2 & 2\omega & 2\omega^2 & 0& 2\omega e^{\frac{13}{18}2\pi i} & 2\omega^2 e^{\frac{13}{18}2\pi i} & 2 e^{\frac{13}{18}2\pi i} & 2\omega^2 e^{-\frac{13}{18}2\pi i} & 2\omega e^{-\frac{13}{18}2\pi i} &2 e^{-\frac{13}{18}2\pi i}\\
2 & 2\omega & 2\omega^2 & 0& 2\omega^2 e^{\frac{13}{18}2\pi i} & 2 e^{\frac{13}{18}2\pi i} & 2\omega e^{\frac{13}{18}2\pi i} & 2\omega e^{-\frac{13}{18}2\pi i} & 2 e^{-\frac{13}{18}2\pi i} &2\omega^2 e^{-\frac{13}{18}2\pi i}\\
\end{pmatrix}
\label{Z3SO8Smatrix}\end{align}
where $\mathcal{D}=6$.
\end{widetext}
The fusion algebra can be obtained by using the Verlinde formula in Eq.\eqref{verlinde}: 
\begin{align}
&
z_3\times z_3=\overline{z}_3,\qquad z_3\times\overline{z}_3=1,
\nonumber\\
&
\rho_n\times z_3=\rho_{n+1},\qquad \rho_n\times\overline{z}_3=\rho_{n-1},
\nonumber\\
&
z_3\times\psi=\overline{z}_3\times\psi=\psi,
\nonumber\\
&
\rho_n\times\rho_n=\overline{\rho}_{n-1}+\overline{\rho}_n,
\nonumber\\
&
\psi\times\rho_n=\rho_0+\rho_1+\rho_2,
\nonumber\\
&
\psi\times\psi =1+z_3+\overline{z}_3+2\psi, 
\end{align}
where $n=0,1,2$.

The $SO(8)_1/\mathbb{Z}_3$ orbifold CFT is very similar to $SU(3)_3$ CFT.\cite{Alimohammadi94,BarkeshliBondersonChengWang14,TeoHughesFradkin15} The chiral $SU(3)_3$ CFT has ten primary fields and they are labelled by the dimensions of the truncated irreducible representation of $su(3)$. Their conformal dimensions and quantum dimensions are listed in Table \ref{su3_3}. We can stack a $\mathbb{Z}_3$ SPT phase discussed in Sec.~\ref{orb_z3_spt} on top of $SO(8)_1$ topological phase and gauge $\mathbb{Z}_3$ symmetry for the composite system. Since the cohomology class for $\mathbb{Z}_3$ group is $H^3(\mathbb{Z}_3,U(1))=\mathbb{Z}_3$, we can obtain three different $\mathcal{S}$ and $\mathcal{T}$ matrices.\cite{BarkeshliBondersonChengWang14,TeoHughesFradkin15}

\begin{table}[t]
  \begin{center}
\begin{ruledtabular}
    \begin{tabular}{c| c c c c c c c c c c}
      $SU(3)_3$ & 1 & \bf{3} & $\overline{\bf 3}$ & \bf{6} & $\overline{\bf 6}$ & \bf{8} & \bf{10} & $\overline{\bf 10}$ & \bf{15} & $\overline{\bf 15}$\\
      \hline
	$SO(8)_1/\mathbb{Z}_3$ & $0\bar{0}$ & $1\bar{1}$ & $2\bar{2}$ & $1\bar{2}$ & $2\bar{0}$ & $0\bar{1}$ & $2\bar{1}$ & $0\bar{2}$ & $1\bar{0}$& 1\\
$h_{\chi}$ & 0 & $\frac{2}{9}$ & $\frac{2}{9}$ & $\frac{5}{9}$ & $\frac{5}{9}$ & $\frac{1}{2}$ & 1 & 1 & $\frac{8}{9}$ & $\frac{8}{9}$\\
$d_{\chi}$ & 1 & 2 & 2 & 2 & 2 & 3 & 1 & 1 & 2 & 2\\ 
    \end{tabular}
    \end{ruledtabular}
 \caption{
The first line is the quasiparticle excitation for $SU(3)_3$ topological phase. The second line is the corresponding quasiparticle excitation for $SO(8)_1/\mathbb{Z}_3$ topological phase. The third and fourth lines are conformal dimensions $h_\chi$ and quantum dimensions $d_\chi$ for them.
    \label{su3_3}
}
  \end{center}
\end{table}

\subsubsection{Orbifolding $\mathbb{Z}_2$ symmetry}

The reflection symmetry operator $\sigma_1$ is defined as
\begin{equation}
\sigma_{1}=\begin{pmatrix} 1&0&0&0\\0&1&0&0\\0&0&0&1\\0&0&1&0\end{pmatrix}
\end{equation}
Under this symmetry, $\psi_{\lambda_1}$ is invariant while $\psi_{\lambda_2}$ and $\psi_{\lambda_3}$ are exchanged. This means that after gauging the reflection symmetry $\sigma_1$, $\mathbb{Z}_2$ flux will have species labels and are denoted as $\Sigma^0$ and $\Sigma^1$. The $\mathbb{Z}_2$ flux can be attached by the $\mathbb{Z}_2$ charge to form the composite particle. $\psi_{\lambda_2}$ and $\psi_{\lambda_3}$ will group together to form the super-selection sector $\psi$. The $\mathbb{Z}_2$ invariant anyon $\psi_{\lambda_0}$ and $\psi_{\lambda_1}$ will both split into two characters, $I$, $z_2$ and $\lambda_1$ and $z_2\lambda_1$. They are differed by a $\mathbb{Z}_2$ gauge charge $z_2$. 

For the boundary CFT, the calculation is analogous to the previous several examples on $\mathbb{Z}_2$ orbifold CFT. Here we will first consider the partitoin function with twist boundary condition in the time direction. Under the $\sigma_{1}$ operator, only the state $\sigma_{1}|\textbf{K}\vec{\Lambda}+\vec{\lambda}\rangle=|\textbf{K}\vec{\Lambda}+\vec{\lambda}\rangle$ will remain. This requires $\vec{\Lambda}^T=(n_1,n_2,n_3,n_3)$ and $\vec{\lambda}=\vec{\lambda}_0, \vec{\lambda}_1$. The partition function with twist boundary condition in the time direction equals to
\begin{align}
\mathcal{Z}^{0,\frac{1}{2}}=\left(\frac{1}{\eta}\right)^2\sum_{\vec{\Lambda}}q^{\frac{1}{2}(\vec{\Lambda}^T+\vec{\lambda}_j^T)\textbf{K}(\vec{\Lambda}+\vec{\lambda}_i)}\sqrt{\frac{\eta}{\theta_2}}
=C_j\sqrt{\frac{\eta}{\theta_2}}
\end{align}
$C_j$ is the characters for the chiral CFT with the $\mathrm{K}$-matrix
\begin{equation}
\textbf{K}_{z_2}=\begin{pmatrix} 2 & -1&-2 \\ -1&2&0\\-2 &0&4\end{pmatrix}
\end{equation}
after projecting to the three-dimensional $\mathbb{Z}_2$ symmetric lattice $\Lambda=(n_1,n_2,n_3,n_3)$.
This $\mathrm{K}$-matrix has four quasi-particles $c_0^T=(0,0,0),c_1^T=(0,1,0), c_2^T=(0,0,1), c_3^T=(0,1,1)$. For $\mathcal{Z}^{0\frac{1}{2}}$, $j$ can only take 0 or 1. The four characters have conformal weight $h_{c_0}=1, h_{c_1}=-1, h_{c_2}=e^{2\pi i\frac{3}{8}}, h_{c_3}=e^{2\pi i\frac{3}{8}}$.
The $\mathcal{S}$-matrix for this $\mathbf{K}_{z_2}$ matrix is
\begin{equation}
\mathcal{S}=\frac{1}{2}\begin{pmatrix} 1 & 1&1 &1 \\ 1&1&-1&-1\\1 &-1&i&-i\\1&-1&-i&i\end{pmatrix}.
\end{equation}
All the other sectors $\mathcal{Z}^{\mu\nu}$ ($\mu,\nu=0,\frac{1}{2}$) can be obtained by performing a series of $S$ and $T$ transformation on $\mathcal{Z}^{0,\frac{1}{2}}$.

The characters for $SO(8)_1/\mathbb{Z}_2$ CFT are listed in the Table. \ref{so8z2}:
\begin{table}[t]
\centering
\begin{ruledtabular}
\begin{tabular}{lll}
character $\chi$ & $d_\chi$ & $h_\chi$ \\\hline
$ \chi_I=\chi_{\lambda_0}+C_0\sqrt{\frac{\eta}{\theta_2}}$ & $1$ & $0$\\
$\chi_{z_2}=\chi_{\lambda_0}-C_0\sqrt{\frac{\eta}{\theta_2}}$ & $1$ & $0$\\
$\chi_{\lambda_1}=\chi_{\lambda_1}+C_1\sqrt{\frac{\eta}{\theta_2}}$ & $1$ & $1/2$\\
$\chi_{z_2\lambda_1}=\chi_{\lambda_1}-C_1\sqrt{\frac{\eta}{\theta_2}}$ & $1$  & $1/2$\\
$\chi_{\psi}=\chi_{\lambda_2}+\chi_{\lambda_3}$ & $2$ & $1/2$\\
$\chi_{\Sigma^0}=\frac{1}{\sqrt{2}}(C_0+C_1)\sqrt{\frac{\eta}{\theta_4}}+\frac{1}{\sqrt{2}}(C_0-C_1)\sqrt{\frac{\eta}{\theta_3}}$ & $\sqrt{2}$ & $1/16$\\
$\chi_{z_2\Sigma^0}=\frac{1}{\sqrt{2}}(C_0+C_1)\sqrt{\frac{\eta}{\theta_4}}-\frac{1}{\sqrt{2}}(C_0-C_1)\sqrt{\frac{\eta}{\theta_3}}$ & $\sqrt{2}$ & $9/16$\\
$\chi_{\Sigma^1}=\frac{1}{\sqrt{2}}(C_2+C_3)\sqrt{\frac{\eta}{\theta_4}}+\frac{1}{\sqrt{2}}(C_2+C_3)\sqrt{\frac{\eta}{\theta_3}}$ & $\sqrt{2}$ & $7/16$\\
$\chi_{z_2\Sigma^1}=\frac{1}{\sqrt{2}}(C_2+C_3)\sqrt{\frac{\eta}{\theta_4}}-\frac{1}{\sqrt{2}}(C_2+C_3)\sqrt{\frac{\eta}{\theta_3}}$ & $\sqrt{2}$ & $15/16$
\label{so8z2}
\end{tabular}
\end{ruledtabular}
\caption{The quantum dimensions $d_\chi$ and spin statistics $\theta_\chi=e^{2\pi ih_\chi}$ of deconfined fluxes, charges and super-sectors from orbifolding the $\mathbb{Z}_2$ symmetry of $SO(8)_1$. 
}
\end{table}

The $\mathcal{S}$ matrix for this model is
\begin{equation}
\mathcal{S}=\frac{1}{4}\left(\begin{smallmatrix}1&1&1&1&2&\sqrt{2}&\sqrt{2}&\sqrt{2}&\sqrt{2}\\
1&1&1&1&2&-\sqrt{2}&-\sqrt{2}&-\sqrt{2}&-\sqrt{2}\\
1&1&1&1&-2&\sqrt{2}&\sqrt{2}&-\sqrt{2}&-\sqrt{2}\\
1&1&1&1&-2&-\sqrt{2}&-\sqrt{2}&\sqrt{2}&\sqrt{2}\\
2&2&-2&-2&0&0&0&0&0\\
\sqrt{2}&-\sqrt{2}&\sqrt{2}&-\sqrt{2}&0&0&0&2&-2\\
\sqrt{2}&-\sqrt{2}&\sqrt{2}&-\sqrt{2}&0&0&0&-2&2\\
\sqrt{2}&-\sqrt{2}&-\sqrt{2}&\sqrt{2}&0&2&-2&0&0\\
\sqrt{2}&-\sqrt{2}&-\sqrt{2}&\sqrt{2}&0&-2&2&0&0
\end{smallmatrix}\right)
\end{equation}
This theory has identical topological content as $SO(1)_1 \times SO(7)_1$, where $SO(1)_1$ stands for the usual chiral Ising theory with $h_\sigma = 1/16$.\cite{TeoHughesFradkin15} Following the same method we used in Sec.~\ref{SPT_SET}, we can further stack a $\mathbb{Z}_2$ SPT phase on top of $SO(8)_1$ and then gauge $\mathbb{Z}_2$ symmetry for the composite system. The orbifold CFT on the boundary is equivalent to $SO(3)_1 \times SO(5)_1$ CFT. Since chiral $SO(n)_1$ CFT has central charge $c=n/2$, the total central charge for $SO(n)_1 \times SO(8-n)_1$ CFT is 4 and is the same as the central charge for $SO(8)_1$ CFT before gauging.

\subsubsection{Orbifolding $S_3$ symmetry of $SO(8)_1$ CFT}
At last, we will orbifold the full $S_3$ symmetry for $SO(8)_1$ state. As we discussed before, the $SO(8)_1/\mathbb{Z}_3$ orbifold CFT is a $SU(3)_3$-like CFT. It has ten characters including three $\mathbb{Z}_3$ charges $1$, $z_3$ and $\overline{z}_3$, six threefold fluxes $\rho_n$ and $\overline{\rho}_n$ and a superselection sector. Since $S_3=\mathbb{Z}_2\ltimes\mathbb{Z}_3$, after orbifolding $\mathbb{Z}_3$ symmetry in $SO(8)_1$ CFT, there is still a remaining $\mathbb{Z}_2$ symmetry, which switches $z_3$ and $\overline{z}_3$ as well as $\rho_n$ and $\overline{\rho}_n$. Orbifolding the full $S_3$ symmetry of $SO(8)_1$ is therefore equivalent to orbifolding the $\mathbb{Z}_2$ conjugation symmetry of the $SU(3)_3$-like state.\cite{TeoHughesFradkin15,BarkeshliBondersonChengWang14} 

After orbifolding $\mathbb{Z}_2$ symmetry, the non-self-conjugate characters will combine together and group into super-selection sectors, which include threefold charge $[z_3+\overline{z}_3]$ and fluxes $[\rho_0+\overline{\rho}_0]$. The composite flux-charge particle $[\rho_i+\overline{\rho}_i]$ ($i=1,2$) are differed from $[\rho_0+\overline{\rho}_0]$ by $\mathbb{Z}_3$ charges. The vacuum will again split into two sectors which are differed by a $\mathbb{Z}_2$ charge. Their characters can be calculated by applying projection operator defined in Eq.\eqref{char_def} on the vacuum character of $SU(8)_1/\mathbb{Z}_3$.  The original superselection sector $[\psi_1+\psi_2+\psi_3]$ will also carry $\mathbb{Z}_2$ charge and their characters can also be obtained by applying $\mathbb{Z}_2$ projection operator on it. Finally, we can calculate the characters for the $\mathbb{Z}_2$ twist fields. They have species labels and carry $\mathbb{Z}_2$ charge. Actually, their explicit form is the same the characters for the $SO(8)_1/\mathbb{Z}_2$ orbifold CFT but with different quantum dimensions.

The Tables. \ref{so8s3} is the characters for the orbifolding $S_3$ symmetry of $SO(8)_1$ theory

\begin{table}[t]
\centering
\begin{ruledtabular}
\begin{tabular}{lll}
character $\chi$ & $d_\chi$ & $h_\chi$ \\\hline
$\chi_I=\chi_{\lambda_0}+e^{\frac{2\pi i}{12}}\mathcal{Z}^{0,\frac{1}{3}}+e^{-\frac{2\pi i}{12}}\mathcal{Z}^{0,\frac{2}{3}}+C_0\sqrt{\frac{\eta}{\theta_2}}$ & $1$ & $0$\\
$\chi_{z_2}=\chi_{\lambda_0}+e^{\frac{2\pi i}{12}}\mathcal{Z}^{0,\frac{1}{3}}+e^{-\frac{2\pi i}{12}}\mathcal{Z}^{0,\frac{2}{3}}-C_0\sqrt{\frac{\eta}{\theta_2}}$ & $1$ & $0$\\
$\chi_{z_3+\overline{z}_3}=2\chi_{\lambda_0}-e^{\frac{2\pi i}{12}}\mathcal{Z}^{0,\frac{1}{3}}-e^{-\frac{2\pi i}{12}}\mathcal{Z}^{0,\frac{2}{3}}$ & $2$ & $0$\\
$\chi_{\rho_0+\overline{\rho}_0}=\chi_{\rho_0}+\chi_{\overline{\rho_0}}$ & 4 & $1/9$\\
$\chi_{\rho_1+\overline{\rho}_1}=\chi_{\rho_1}+\chi_{\overline{\rho_1}}$ & 4 & $4/9$\\
$\chi_{\rho_2+\overline{\rho}_2}=\chi_{\rho_2}+\chi_{\overline{\rho_2}}$ & 4 & $7/9$\\
$\chi_{\psi}=\chi_{\lambda_1}+\chi_{\lambda_2}+\chi_{\lambda_3}-C_1\sqrt{\frac{\eta}{\theta_2}}$ & $3$ & $1/2$\\
$\chi_{z_2\psi}=\chi_{\lambda_1}+\chi_{\lambda_2}+\chi_{\lambda_3}+C_1\sqrt{\frac{\eta}{\theta_2}}$ & $3$ & $1/2$\\
$\chi_{\Sigma^0}=\frac{1}{\sqrt{2}}(C_0+C_1)\sqrt{\frac{\eta}{\theta_4}}+\frac{1}{\sqrt{2}}(C_0-C_1)\sqrt{\frac{\eta}{\theta_3}}$ & $3\sqrt{2}$ & $1/16$\\
$\chi_{z_2\Sigma^0}=\frac{1}{\sqrt{2}}(C_0+C_1)\sqrt{\frac{\eta}{\theta_4}}-\frac{1}{\sqrt{2}}(C_0-C_1)\sqrt{\frac{\eta}{\theta_3}}$ & $3\sqrt{2}$ & $9/16$\\
$\chi_{\Sigma^1}=\frac{1}{\sqrt{2}}(C_2+C_3)\sqrt{\frac{\eta}{\theta_4}}+\frac{1}{\sqrt{2}}(C_2+C_3)\sqrt{\frac{\eta}{\theta_3}}$ & $3\sqrt{2}$ & $7/16$\\
$\chi_{z_2\Sigma^1}=\frac{1}{\sqrt{2}}(C_2+C_3)\sqrt{\frac{\eta}{\theta_4}}-\frac{1}{\sqrt{2}}(C_2+C_3)\sqrt{\frac{\eta}{\theta_3}}$ & $3\sqrt{2}$ & $15/16$\\
\end{tabular}
\end{ruledtabular}
\caption{The quantum dimensions $d_\chi$ and spin statistics $\theta_\chi=e^{2\pi ih_\chi}$ of deconfined fluxes, charges and super-sectors from gauging the $S_3$ symmetry of $so(8)_1$. 
}\label{so8s3}
\end{table}
\begin{widetext}
The $\mathcal{S}$ matrix is given by 
\begin{align}
\mathcal{S}=\frac{1}{\mathcal{D}}\left(
\begin{smallmatrix}
 1 & 1 & 2 & 4 & 4 & 4 & 3 & 3 & 3 \sqrt{2} & 3 \sqrt{2} & 3 \sqrt{2} & 3 \sqrt{2} \\
 1 & 1 & 2 & 4 & 4 & 4 & 3 & 3 & -3 \sqrt{2} & -3 \sqrt{2} & -3 \sqrt{2} & -3 \sqrt{2} \\
 2 & 2 & 4 & -4 & -4 & -4 & 6 & 6 & 0 & 0 & 0 & 0 \\
 4 & 4 & -4 & -8\cos\frac{4\pi}{9} & -8\cos\frac{8\pi}{9} & -8\cos\frac{16\pi}{9} & 0 & 0 & 0 & 0 & 0 & 0 \\
 4 & 4 & -4 & -8\cos\frac{8\pi}{9} & -8\cos\frac{16\pi}{9} & -8\cos\frac{4\pi}{9} & 0 & 0 & 0 & 0 & 0 & 0 \\
 4 & 4 & -4 & -8\cos\frac{16\pi}{9} & -8\cos\frac{4\pi}{9} & -8\cos\frac{8\pi}{9} & 0 & 0 & 0 & 0 & 0 & 0 \\
 3 & 3 & 6 & 0 & 0 & 0 & -3 & -3 & -3 \sqrt{2} & -3 \sqrt{2} & 3 \sqrt{2} & 3 \sqrt{2} \\
 3 & 3 & 6 & 0 & 0 & 0 & -3 & -3 & 3 \sqrt{2} & 3 \sqrt{2} & -3 \sqrt{2} & -3 \sqrt{2} \\
 3 \sqrt{2} & -3 \sqrt{2} & 0 & 0 & 0 & 0 & -3 \sqrt{2} & 3 \sqrt{2} & 0 & 0 & 6 & -6 \\
 3 \sqrt{2} & -3 \sqrt{2} & 0 & 0 & 0 & 0 & -3 \sqrt{2} & 3 \sqrt{2} & 0 & 0 & -6 & 6 \\
 3 \sqrt{2} & -3 \sqrt{2} & 0 & 0 & 0 & 0 & 3 \sqrt{2} & -3 \sqrt{2} & 6 & -6 & 0 & 0 \\
 3 \sqrt{2} & -3 \sqrt{2} & 0 & 0 & 0 & 0 & 3 \sqrt{2} & -3 \sqrt{2} & -6 & 6 & 0 & 0 
\end{smallmatrix}
\right)\label{S3Smatrix}
\end{align}
where the total quantum dimension is $\mathcal{D}=12$, and the entries are arranged to have the same anyon order in table~\ref{so8s3}. 
\end{widetext}
The fusion algebras for this twist liquid are
\begin{align}
&
z_2\times z_2=1,\quad z_2\times [z_3+\overline{z}_3]=[z_3+\overline{z}_3]\nonumber\\
&
[z_3+\overline{z}_3]\times [z_3+\overline{z}_3]=1+z_2+[z_3+\overline{z}_3]\nonumber\\
&
z_2\times [\rho_n+\overline{\rho}_n]= [\rho_n+\overline{\rho}_n]\nonumber\\
&
z_2\times \psi=z_2\psi,\quad  [z_3+\overline{z}_3]\times\psi=\psi+z_2\psi\nonumber\\
&
z_2\times\Sigma^{\lambda}=z_2\Sigma^{\lambda}\quad  [z_3+\overline{z}_3]\times \Sigma^{\lambda}=\Sigma^{\lambda}+z_2\Sigma^{\lambda}\nonumber\\
&
\Sigma^{\lambda}\times\Sigma^{\lambda}=1+[z_3+\overline{z}_3]+\psi+\sum_{n=0}^2[\rho_n+\overline{\rho}_n]\nonumber\\
&
[\rho_n+\overline{\rho}_n]\times\Sigma^{\lambda}=\Sigma^{0}+z_2\Sigma^{0}+\Sigma^{1}+z_2\Sigma^{1}
\end{align} 
where $n=0,1,2$ and $\lambda=0,1$.

This $\mathcal{S}$ matrix can be modified if we stack SPT phase on top of $SO(8)_1$ topological phase and gauge $S_3$ symmetry for the composite system. In total, we can obtain six different $\mathcal{S}$ matrices, which can be labelled by cohomology classes $ H^3(S_3,U(1))=\mathbb{Z}_3\times\mathbb{Z}_2$.\cite{TeoHughesFradkin15} These six twist liquids also have different $\mathcal{T}$ matrices and fusion rules. The calculation is similar to that in Sec.~\ref{SPT_SET} and will be omitted here. 

\section{Discussion and conclusion}
\label{conclusion}
Topological phases are commonly equipped with some global discrete anyonic symmetry $\mathcal{G}$. For instance, all topological phase carry a conjugation symmetry ${\bf a}\leftrightarrow\overline{\bf a}$ that swaps between anti-partners. Abelian discrete gauge theoies carry electric-magnetic symmetry. Fractional quantum Hall states can be bilayer or multi-layer symmetric. Some topological phases even carry non-abelian global symmetries. After gauging an anyonic symmetry $\mathcal{G}$ in a parent phase, a new topological phase emerges, and is called a  ``twist liquid". The gauge charge, gauge flux, and the anyon in the parent topological phase will manifest in a non-trivial way as quantum quasi-particle excitations in the twist liquid. In our paper, instead of directly studying the bulk $(2+1)$d topological properties for these quasi-particles, we use bulk-boundary correspondence to study the edge $(1+1)$d CFT after orbifolding symmetry $\mathcal{G}$. By constructing the characters for the orbifold CFT, we can calculate the modular $\mathcal{S}$ and $\mathcal{T}$ matrices for the CFT, which are equivalent to that for the bulk topological phase and are essential to understand the bulk topological order.

Gauging symmetry in the bulk corresponds to orbifolding symmetry on the boundary.
In this paper, we first investigate SPT phases protected by discrete symmetries and
construct the characters of the orbifold CFTs on the boundary after gauging the symmetries.
We use this method to study the gauged topological phases and identify the
non-trivial SPT phases by analyzing the modular transformation
for the orbifold partition functions.
We further extend this method to long-range entangled topological phases and
explicitly construct the characters of the orbifold CFTs
after gauging anyonic symmetry in the bulk, and study the new topological phases
after gauging anyonic symmetries.
In particular, we demonstrate in the $D(\mathbb{Z}_N)$ quantum double model,
the chiral $SU(3)_1$ phase and the chiral $SO(8)_1$ topological phases.
These three examples are globally symmetric parent phases that carry abelian
topological orders.
On the other hand, we can also start with a non-abelian parent phase and gauge
its anyonic symmetry.
For instance, the critical 4-state Potts model is described by
$SU(2)_1/Dih_2=U(1)_4/\mathbb{Z}_2$ CFT,
where $Dih_2$ is the dihedral group embedded in $SU(2)$ that contains
$\pi$-rotations about the $x$-, $y$- and $z$-axes.\cite{bigyellowbook,
  Ginsparg88, DijkgraafVafaVerlindeVerlinde99, CappelliAppollonio02,
  TeoHughesFradkin15}
The non-abelian topological phase with chiral 4-state Potts CFT as the boundary
has the anyonic symmetry group $S_3=\mathbb{Z}_2\ltimes\mathbb{Z}_3$.
Similar to what we did for the $SO(8)_1$ CFT,
taking the $\mathbb{Z}_2$, $\mathbb{Z}_3$ and $S_3$ orbifolds of
$SU(2)_1/Dih_2$ leads to a series of topological twist liquids.
\begin{align}\vcenter{\hbox{\includegraphics[width=0.3\textwidth]{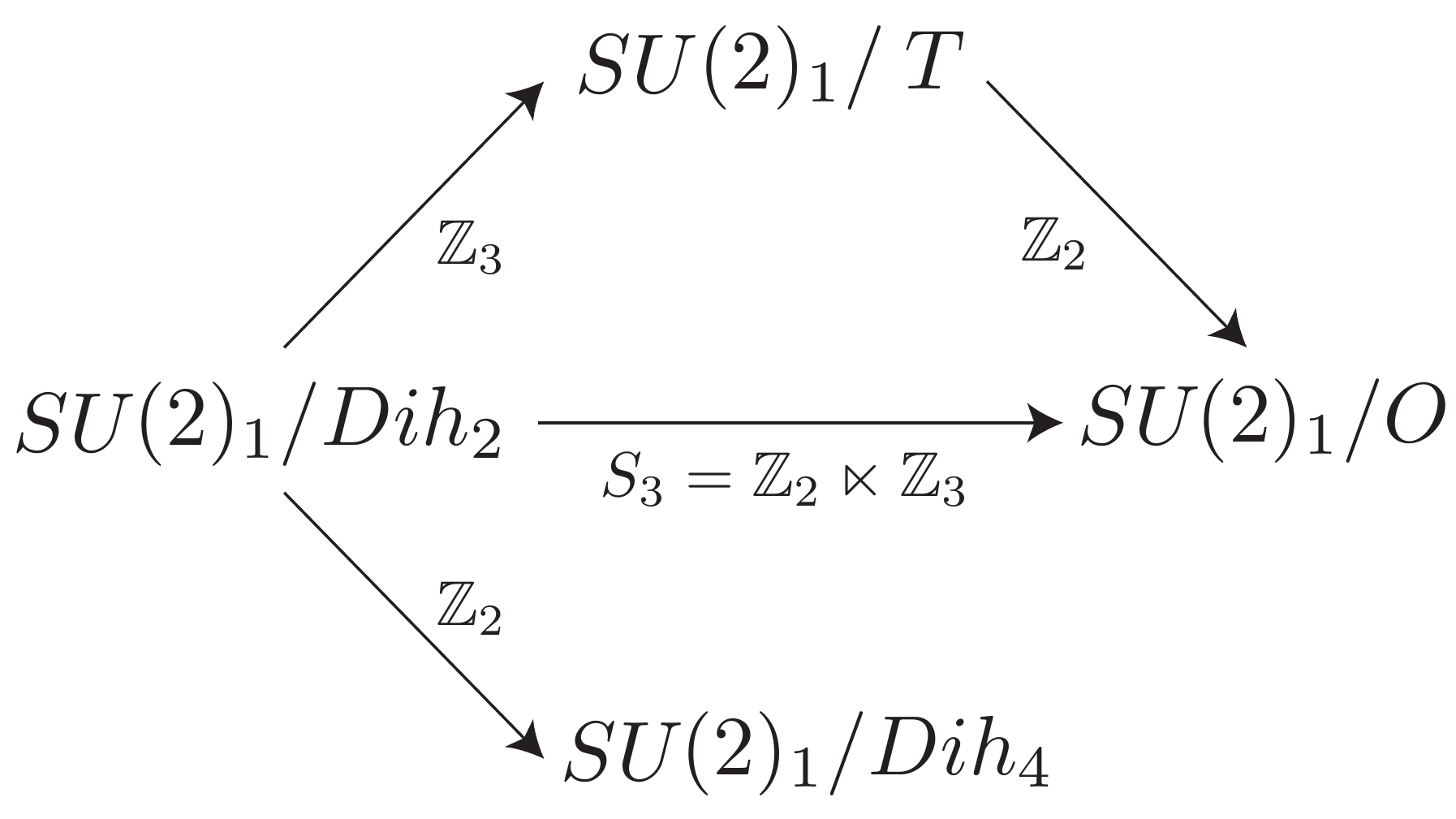}}}\label{eq:4statePottsgaugingdiagram}\end{align}
where $T$ is the tetrahedron group and $O$ is the octahedron group. The details
for these twist liquids and orbifold CFTs were discussed
in Ref.\ \onlinecite{CappelliAppollonio02, DijkgraafVafaVerlindeVerlinde99, Ginsparg88, TeoHughesFradkin15}.

Finally, in this paper, we focus on the orbifold CFTs for bosonic topological
phases in 2+1 dimensions.
This set the stage for investigation of twist liquids in higher dimensions as well as an extension to fermionic systems.\cite{3d-prep}

\acknowledgements
We thank Eduardo Fradkin for useful discussions. We thank the organizers of the KITP program “Symmetry,
Topology, and Quantum Phases of Matter: From Tensor
Networks to Physical Realizations”. This work is supported by the NSF under Grant No.\ DMR-1455296 (SR), DMR-1653535 (JCYT) and  DMR-1408713 (XC).
XC is also supported by a postdoctoral fellowship from the Gordon and Betty Moore Foundation, under the EPiQS initiative, Grant GBMF4304, at the Kavli Institute for Theoretical Physics. 
AR was supported by the German Research Foundation (DFG) through grants ZI 513/2-1 and HE 7267/1-1.

\bibliography{refFinal}

\end{document}